\documentclass[11pt]{article}

\usepackage[T1]{fontenc}
\usepackage[utf8]{inputenc}

\usepackage{lmodern}
\usepackage{euler}
\usepackage{microtype}

\usepackage[margin=1.125in]{geometry}
\usepackage{setspace}
\usepackage{amsmath,amssymb,amsthm}
\usepackage{mathtools}
\usepackage{bm}

\usepackage{tikzpagenodes}
\usepackage{background}
\backgroundsetup{
  scale=5,                
  color=gray,             
  opacity=0.08,           
  angle=45,               
  contents={\Large REVISED VERSION}, 
  position=current page.center,  
  vshift=0.5cm             
}

\usepackage{graphicx}
\usepackage{booktabs}
\usepackage{threeparttable}
\usepackage{longtable}
\usepackage{multirow}
\usepackage{mathpazo}

\usepackage{subcaption}
\usepackage{array}
\usepackage{amssymb}
\usepackage{pifont}

\usepackage[numbers,sort&compress]{natbib}

\usepackage[
colorlinks=true,
linkcolor=blue,
citecolor=blue,
urlcolor=blue
]{hyperref}

\usepackage[nameinlink]{cleveref}

\title{\bfseries
Strategic Effort and Non-Linear Positional Bandwagon Drafting Benefits
in Multi-Stage Competitive Games:\\
Evidence from Triathlon}

\author{
Felix Reichel\thanks{\textit{Corresponding author: F.Reichel \url{kontakt<at>felixreichel<dot>com}}}
}

\date{July/August 2026}

\begin{document}

\maketitle

\begin{center}
\small
\textbf{\textit{TBP}}\\
\large Sport Economics Research \textbf{3}(1)
\end{center}

\begin{abstract}

This paper examines strategic effort and positioning choices in finite
multi-stage games. These choices can generate positional bandwagon
drafting benefits through externalities when athletes follow
others\textquotesingle{} trajectories. Focusing on open-water swim
drafting, where athletes reduce drag by swimming directly behind peers
its performance effects on final race outcomes are estimated through an
estimated structural framework with endogenous positioning choices and
predetermined effort. Leveraging exogenous variation from COVID-19
drafting bans in Austrian triathlons, which altered start procedures and
disrupted standard group formations, a panel leave-one-out group ability
instrumental variables strategy is applied to isolate the causal
nonlinear effect of positional drafting. Restricted-sample and pooled
positional drafting benefits IV estimates reveal non-linear concave
gains: in small groups each deeper drafting position improves finishing
rank by over 30\%, with diminishing returns in larger groups.

\end{abstract}


\smallskip


\newpage

\doublespacing
\section{Introduction}

Drafting plays a vital role in race dynamics and performance outcomes in
endurance and multi-disciplinary sports (Chatard et al., 1990; Chatard
\& Wilson, 2003). In triathlon, where races consist of three finite
stages---namely swimming, cycling, and running---drafting introduces a
strategic externality that vastly varies by discipline, race format and
finally race regulation (Leitner, 2021; Tullock, 1980). While drafting
in professional cycling is the most visible and widely studied form of
it in sport economics, hydrodynamic drag reduction, studied rather
extensively in physiology and medical literature in the swim segment,
can also propagate into succeeding cycling performance (Delextrat et
al., 2003) and influence overall race outcomes (Reichel, 2025).
Open-water triathlon therefore provides a unique empirical setting for
studying athletes\textquotesingle{} drafting behaviour and its
performance implications (Bolon et al., 2023; Reichel, 2025).

Triathlon was selected over other endurance sports for several reasons.
First, it combines sequential disciplines within a single competition,
allowing effects of one stage to propagate through subsequent stages
(Delextrat et al., 2003). Second, substantial variation in race formats,
start policies, and field sizes provides natural variation in drafting
group formation (Leitner, 2021; Reichel, 2025). Third, unlike
single-discipline sports where such variation on a similar scale is
limited (Chatard \& Wilson, 2003; Bolon et al., 2023), triathlon offers
a saturated panel structure with repeated athlete observations across
multiple events and time periods.

Empirical estimates of drafting effects in the swim segment under real
race environments remain very limited due to various measurement
challenges. Group formations are often endogenous and cannot be observed
directly, and drafting exposure may correlate with unobserved ability
through ability sorting and selection. Laboratory and biomechanical
studies, including Chatard (1990, 2003), Delextrat (2003), and Bolon et
al. (2023), highlight the energy and performance effects of drafting,
but large-scale causal evidence from race data remains largely
understudied (Reichel, 2025). In this respect, the COVID-19 drafting
bans implemented in Austrian triathlons provide a useful natural
experiment, as they altered start procedures and disrupted default
mostly non-random standard group formations (elitism), generating random
exogenous variation in drafting exposure, which is unlikely to vary completely
over the first race stage.

In the theoretical framework, the standard Tullock contest model is
extended and incorporates strategic externalities to develop a
structural model in which athletes are assumed to first choose swimming
effort and then endogenously select drafting position (Tullock, 1980).
Drafting benefits are modelled with a nonlinear function that implies
diminishing marginal returns beyond a positional threshold based on
previous literature (Reichel, 2025), consistent with contest settings
featuring strategic interaction and heterogeneous payoffs (Tullock,
1980; Vives, 1990; Topkis, 1998). This approach also draws partially on
supermodularity theory to characterise complementarity between effort
and positioning choices under strategic constraints (Topkis, 1998;
Vives, 1990; Milgrom \& Roberts, 1990). The model extends typical work
on dynamic game frameworks in competitive environments (Jeong, 2025).
The model accounts for peer effects and externalities within drafting
groups, recognising that individual payoffs depend on both personal
effort and group composition (Angrist, 2014; Bramoullé et al., 2009).
Such identification challenges mirror those encountered in peer effects
research through social networks, where distinguishing true interaction
effects from correlated baseline characteristics requires careful
econometric design (Bramoullé et al., 2009; Moffitt, 2001). These
considerations align with instrumental variable strategies for
addressing endogeneity at local group-level treatments (Imbens \&
Angrist, 1994; Wooldridge, 2010; Angrist \& Pischke, 2009). The
model\textquotesingle s structural implications are then embedded in an
empirical two-stage least squares framework. To address endogeneity in
within-group drafting positions, a leave-one-out average of peer
swimming ability is used as an instrument, following approaches first
established in productivity estimation with control functions and
instrumental variables (De Loecker, 2011; Huntington-Klein, 2023). This
strategy parallels empirical efforts to isolate drafting effects in
open-water triathlon using exogenous policy variation (Reichel, 2025).
The study contributes to the broader literature on multi-stage strategic
behaviour by offering a simple two-stage framework for effort choice
under externality constraints, and it aligns with work on strategic
complementarities and dynamic decision-making in games (Vives, 1990;
Topkis, 1998). It also addresses methodological concerns regarding peer
effect identification and specification (Angrist, 2014). Beyond these
theoretical contributions, the paper makes three substantive advances: (1) it provides the first in-depth large-scale
causal estimates of swim-stage drafting effects using administrative
race data; (2) it demonstrates how institutional variation from
pandemic-era regulations can identify positional externalities in
competitive games; and (3) it offers a replicable framework for studying
strategic positioning in multi-stage competitive structures across
sports and non-sport contexts.

The rest of the paper is organised as follows. Section 2 describes the
data and institutional setting, including race formats, regulations,
transition phases (T1 and T2), and the construction and numerical
justification of drafting group identifiers from swim-out segment times.
Section 3 develops the theoretical model, extending the Tullock contest
framework to incorporate strategic externalities and endogenous drafting
position choice in a two-stage effort game with diminishing marginal
returns. Section 4 outlines the empirical strategy, detailing the
two-stage least squares framework and the leave-one-out peer ability
instrument used to address endogeneity in within-group positioning while
accounting for measurement error. Section 5 reports the main results,
including instrumental variable estimates of drafting gains, evidence of
diminishing marginal returns across group positions, and robustness
checks across restricted samples and alternative peer group clustering
thresholds. Section 6 concludes with a discussion of implications for
competitive game theory and sport economics, acknowledges limitations of
the empirical strategy, and suggests directions for future research in
similar environments.\textbf{\hfill\break
}

\section{Methods}

\emph{Data}

Triathlon offers an ideal empirical setting for studying drafting
behaviour and its performance implications. Unlike cycling, where
drafting has been studied primarily from an aerodynamic perspective,
triathlon combines sequential disciplines---swimming, cycling, and
running---within a single race, allowing the effects of swim-stage
drafting to propagate through subsequent stages and influence overall
race outcomes (Delextrat et al., 2003). Furthermore, the sport exhibits
substantial variation in race formats, start policies, and field sizes,
providing natural variation in drafting group formation (Leitner, 2021;
Reichel, 2025). These features make triathlon uniquely suited to
identifying the causal effects of positional choice under strategic
externalities, in contrast to single-discipline sports where such
variation is limited (Chatard \& Wilson, 2003; Bolon et al., 2023). A
saturated administrative dataset provided by Triathlon Statistics
Austria covering all official triathlon races that took place from 2010
to 2024 is used in this study. The sample includes sprint, short,
middle, and long-distance formats and contains athlete-level information
on swim-out segment times, total race times, and final overall ranks
after T1 (the first transition from swimming to cycling), cycling, T2
(the second transition from cycling to running), and running. The data
are organized into three relational tables: an athlete table with
identifiers, gender, and birth year; an event table with event
identifiers, calendar dates, and event category; and a result table with
swim-out time, total race time, and final rank for each athlete-event
observation. These tables are merged using the respective foreign keys
to construct an athlete-event panel. From this panel, age is derived as
event year minus birth year, along with age squared, and binary
indicators for gender and period membership (pre-COVID, COVID, and
post-COVID) are encoded. Events are assigned to pandemic periods
according to timing and start-policy regime, capturing the staggered
restrictions in 2020 and their subsequent relaxation. The initial
dataset contains 175,740 observations. After excluding missing values,
DNFs (did not finish), and DNSs (did not start), the cleaned panel
includes 168,391 athlete-event observations.

\emph{Drafting Group Construction}

Drafting groups are inferred using a single-linkage hierarchical
agglomerative clustering algorithm applied to swim-out segment times
within each event. Athletes are assigned to the same cluster if their
swim-out times differ by no more than 5 seconds. This threshold is
empirically motivated: wider thresholds of 10 to 15 seconds generated
unstable or overly broad clusters, while 5 seconds provides a more
conservative grouping rule. To make this choice more interpretable, a
5-second gap corresponds to a physically meaningful separation in the
water given typical open-water swim speeds and is consistent with the
narrow range over which hydrodynamic drafting benefits are expected to
be strongest. Concretely, at a typical open-water race pace of roughly
1.2--1.4 m/s, a 5-second gap corresponds to approximately 6--7 meters of
in-water separation, a distance consistent with the drafting zones
identified in swimming hydrodynamics research (Chatard \& Wilson, 2003).
Within each cluster, the fastest athlete is designated as the leader,
and remaining athletes are assigned ordinal positions based on swim-out
time order. Drafting-position measures are constructed as continuous
variables (ranging from 1 for leader to k for last drafter in a group of
size k) alongside binary indicators such as first drafter and last
drafter. This approach captures heterogeneous positional effects while
preserving the within-group ranking structure.

\emph{Instrumental Variable Strategy and Attenuation Bias}

Measurement error in group composition arises when athletes are
misclassified into drafting groups due to timing uncertainty,
common-pace wave-start variations, or unobserved rare re-grouping during
competition. Such misclassification attenuates estimated positional
coefficients toward zero, biasing inference downward. A leave-one-out
instrumental variable---the mean swimming ability of co-racers excluding
the focal athlete---is used to address this endogeneity. This instrument
satisfies the exclusion restriction because (i) peer ability predicts
group-level drafting competitiveness and thus influences an
individual\textquotesingle s relative position, yet (ii) conditional on
race controls, it is orthogonal to idiosyncratic performance shocks
affecting the focal athlete\textquotesingle s outcome. The leave-one-out
construction prevents mechanical correlation between the instrument and
the dependent variable that would arise if the focal
athlete\textquotesingle s own swim time were included in the
peer-measure aggregate. In the second stage, predicted drafting
positions from the first-stage regression identify causal effects net of
attenuation bias, recovering unbiased estimates of positional advantages
in the presence of group misclassification. Measurement concerns and the
exclusion restriction are further addressed in the Empirical Strategy
section, where potential threats from COVID-era policy changes are
explicitly evaluated.

\emph{Descriptive Statistics and Sample Balance Checks}

Summary statistics indicate substantial heterogeneity across events and
athletes. The average swim-out time is reported in seconds, with
standard deviations varying across race formats. Athletes are
predominantly male, and the age distribution is broad, with a median age
that reflects the adult amateur and competitive nature of the sample.
Gender and age distributions are broadly stable across periods, although
a slight aging pattern appears in the post-2020 period.

\includegraphics[width=3.56798in,height=2.34838in]{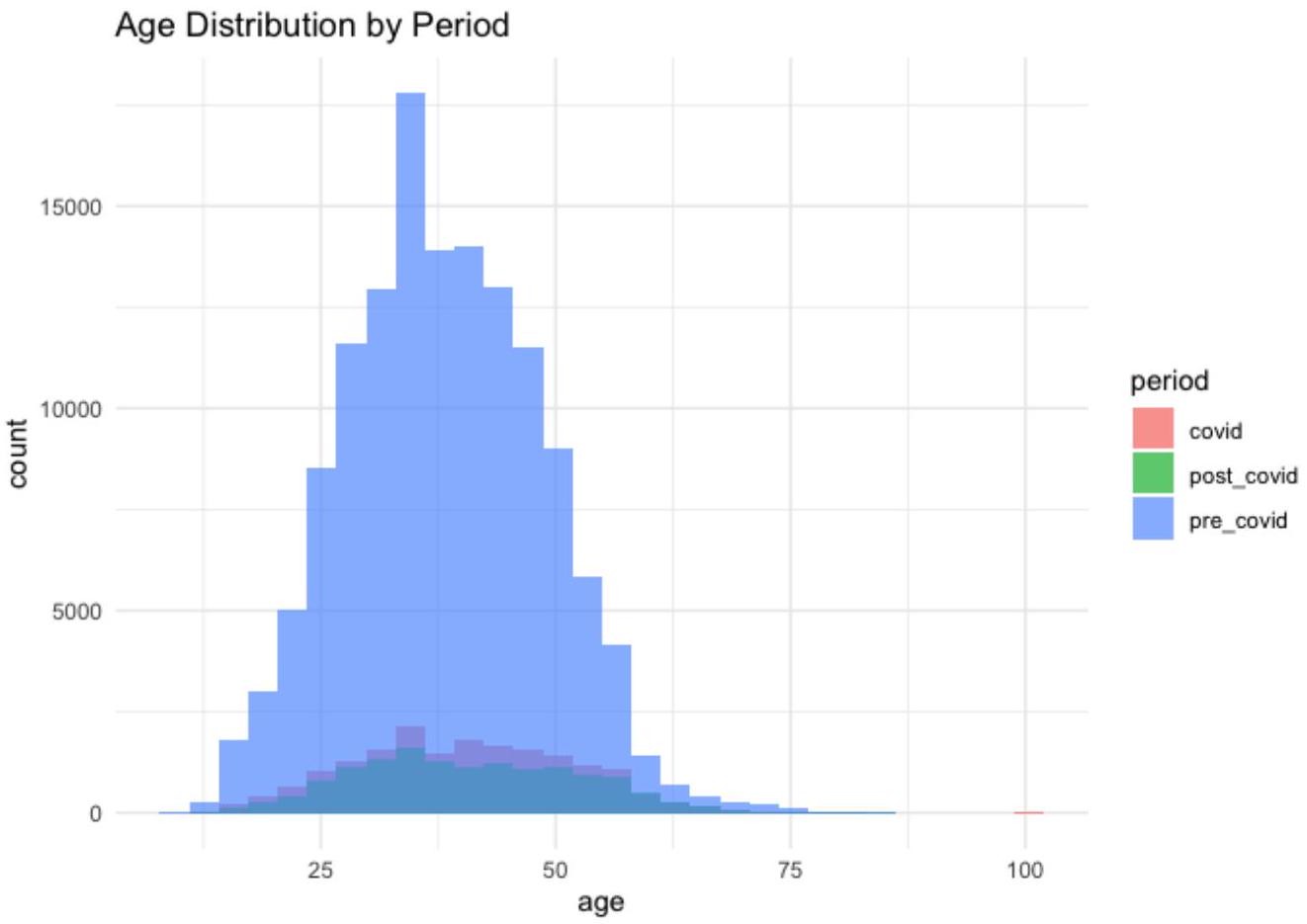}

\textbf{Figure 1.} Age distribution by period balance check

Table 1 compares mean swim-out times across event categories and
pandemic periods. Sprint and short races display higher average swim-out
times after COVID-19 than before, while middle and long-distance events
are comparatively stable.

\includegraphics[width=2.94049in,height=1.87252in]{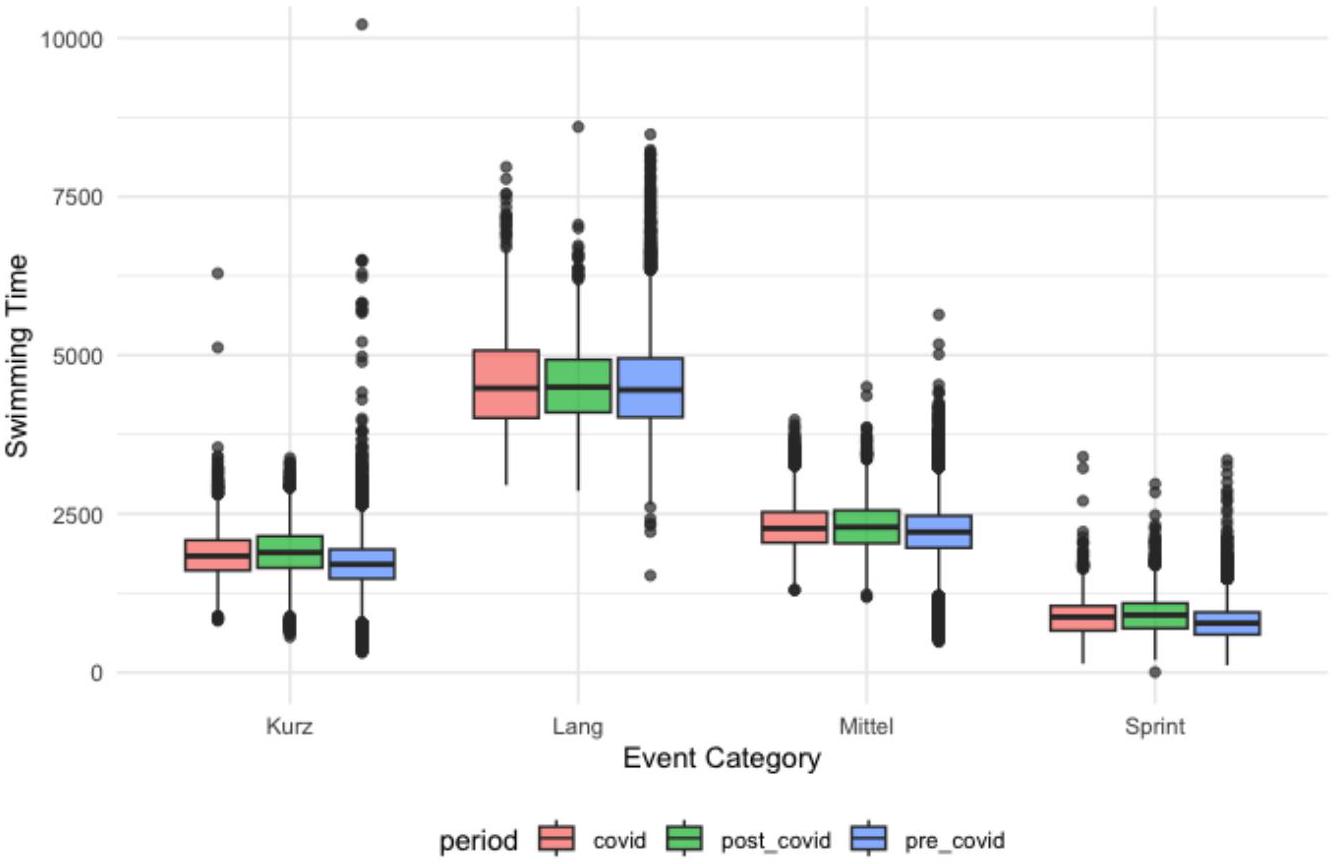}

\textbf{Figure 2.} Swim out times by event category and period balance
check.

Rather than attributing this pattern to a single mechanism, it is more
cautious to note that several factors may contribute, including
pandemic-related detraining, altered start procedures, reduced group
drafting efficiency under staggered starts, and possible changes in
course or timing administration. This broader interpretation is more
consistent with the institutional setting and avoids overstating a
specific mechanism without direct evidence. These factors---particularly
altered start procedures and reduced drafting opportunities---are acknowledged as potential confounders for the exclusion restriction as
defined afterwards and are controlled for partially through event fixed effects
and period interaction terms in the main specification. This setup,
featuring athlete-level panels, inferred swim-group formation, and
quasi-random variation in drafting exposure, provides a suitable
empirical basis for testing the theoretical model developed in the next
section (see also Tables 1--4 and Figs. 1--2 for descriptive statistics
and full balance checks).

\textbf{Table 1.} Mean swim-out times by event category and pandemic
period

\begin{longtable}[]{@{}
  >{\raggedright\arraybackslash}p{(\columnwidth - 8\tabcolsep) * \real{0.1919}}
  >{\raggedright\arraybackslash}p{(\columnwidth - 8\tabcolsep) * \real{0.2383}}
  >{\raggedright\arraybackslash}p{(\columnwidth - 8\tabcolsep) * \real{0.1849}}
  >{\raggedright\arraybackslash}p{(\columnwidth - 8\tabcolsep) * \real{0.1305}}
  >{\raggedright\arraybackslash}p{(\columnwidth - 8\tabcolsep) * \real{0.2544}}@{}}
\toprule()
\begin{minipage}[b]{\linewidth}\raggedright
\textbf{Category}
\end{minipage} & \begin{minipage}[b]{\linewidth}\raggedright
\textbf{Period}
\end{minipage} & \begin{minipage}[b]{\linewidth}\raggedright
\textbf{Mean (s)}
\end{minipage} & \begin{minipage}[b]{\linewidth}\raggedright
\textbf{SD}
\end{minipage} & \begin{minipage}[b]{\linewidth}\raggedright
\textbf{Observations}
\end{minipage} \\
\midrule()
\endhead
\multirow{3}{*}{\textbf{Sprint}} & Pre-COVID & 774 & 258 & 60,871 \\
& COVID & 854 & 295 & 7,985 \\
& Post-COVID & 906 & 296 & 6,152 \\
\multirow{3}{*}{\textbf{Short}} & Pre-COVID & 1,696 & 415 & 37,722 \\
& COVID & 1,855 & 380 & 4,694 \\
& Post-COVID & 1,899 & 403 & 3,726 \\
\multirow{3}{*}{\textbf{Middle}} & Pre-COVID & 2,216 & 436 & 26,711 \\
& COVID & 2,311 & 380 & 4,658 \\
& Post-COVID & 2,315 & 394 & 3,527 \\
\multirow{3}{*}{\textbf{Long}} & Pre-COVID & 4,553 & 773 & 10,201 \\
& COVID & 4,594 & 804 & 1,130 \\
& Post-COVID & 4,572 & 1,035 & 1,014 \\
\bottomrule()
\end{longtable}

\textbf{Table 2.} Summary statistics of numerical variables

\begin{longtable}[]{@{}
  >{\raggedright\arraybackslash}p{(\columnwidth - 12\tabcolsep) * \real{0.2607}}
  >{\raggedright\arraybackslash}p{(\columnwidth - 12\tabcolsep) * \real{0.1099}}
  >{\raggedright\arraybackslash}p{(\columnwidth - 12\tabcolsep) * \real{0.1473}}
  >{\raggedright\arraybackslash}p{(\columnwidth - 12\tabcolsep) * \real{0.1107}}
  >{\raggedright\arraybackslash}p{(\columnwidth - 12\tabcolsep) * \real{0.1099}}
  >{\raggedright\arraybackslash}p{(\columnwidth - 12\tabcolsep) * \real{0.1515}}
  >{\raggedright\arraybackslash}p{(\columnwidth - 12\tabcolsep) * \real{0.1099}}@{}}
\toprule()
\begin{minipage}[b]{\linewidth}\raggedright
\textbf{Variable}
\end{minipage} & \begin{minipage}[b]{\linewidth}\raggedright
\textbf{Min}
\end{minipage} & \begin{minipage}[b]{\linewidth}\raggedright
\textbf{1\textsuperscript{st} Quartile}
\end{minipage} & \begin{minipage}[b]{\linewidth}\raggedright
\textbf{Median}
\end{minipage} & \begin{minipage}[b]{\linewidth}\raggedright
\textbf{Mean}
\end{minipage} & \begin{minipage}[b]{\linewidth}\raggedright
\textbf{3\textsuperscript{rd} Quartile}
\end{minipage} & \begin{minipage}[b]{\linewidth}\raggedright
\textbf{Max}
\end{minipage} \\
\midrule()
\endhead
\textbf{Total} & 165 & 4965 & 8232 & 11789 & 16809 & 61282 \\
\textbf{Swim-Out Times} & 4 & 825 & 1431 & 1625 & 2097 & 29700 \\
\textbf{Race Rank} & 0 & 28 & 69 & 156.9 & 151 & 2623 \\
\textbf{Male} & 0.0000 & 1.0000 & 1.0000 & 0.7898 & 1.0000 & 1.0000 \\
\textbf{Year} & 1922 & 1969 & 1977 & 1977 & 1985 & 2009 \\
\textbf{Covid} & 0.0000 & 0.0000 & 0.0000 & 0.1097 & 0.0000 & 1.0000 \\
\textbf{Post Covid} & 0.00000 & 0.00000 & 0.00000 & 0.08563 & 0.00000 &
1.00000 \\
\textbf{Week Running} & 0.0 & 159.0 & 279.0 & 317.6 & 475.0 & 751.0 \\
\textbf{Age} & 8.00 & 31.00 & 38.00 & 38.56 & 46.00 & 99.00 \\
\textbf{Age Squared} & 64 & 961 & 1444 & 1600 & 2116 & 9801 \\
\textbf{Event Year} & 2010 & 2013 & 2015 & 2016 & 2019 & 2024 \\
\textbf{Cluster (Swim Group)} & 1.00 & 4.00 & 10.00 & 16.95 & 23.00 &
224.00 \\
\textbf{Leader} & 0.0000 & 0.0000 & 0.0000 & 0.3134 & 1.0000 & 1.0000 \\
\textbf{Drafter} & 0.0000 & 0.0000 & 1.0000 & 0.6866 & 1.0000 &
1.0000 \\
\textbf{Drafter Position} & 1.00 & 1.00 & 3.00 & 17.15 & 11.00 &
937.00 \\
\textbf{First Drafter} & 0.0000 & 0.0000 & 0.0000 & 0.2961 & 1.0000 &
1.0000 \\
\textbf{Second Drafter} & 0.0000 & 0.0000 & 0.0000 & 0.1342 & 0.0000 &
1.0000 \\
\textbf{Third Drafter} & 0.00000 & 0.00000 & 0.00000 & 0.08329 & 0.00000
& 1.00000 \\
\textbf{Fourth Drafter} & 0.00000 & 0.00000 & 0.00000 & 0.05846 &
0.00000 & 1.00000 \\
\textbf{Fifth Drafter} & 0.000 & 0.000 & 0.000 & 0.044 & 0.000 &
1.000 \\
\textbf{Last Drafter} & 0.0000 & 0.0000 & 0.0000 & 0.2961 & 1.0000 &
1.0000 \\
\bottomrule()
\end{longtable}

\textbf{Note}: This table presents summary statistics for the key
variables used in the analysis. The dataset consists of 168,391
observations. Variables include athlete characteristics, race details,
and encoded swim group (cluster), drafting, drafting position binary
dummy variables. Encoded categorical variables such as Event Category
Sprint, Short, Middle, Long are not displayed. The variable Week Running
represents weeks races were taking place per year.

\textbf{\hfill\break
Table 3.} Balance check of swim-out times across different event
categories and periods

\begin{longtable}[]{@{}
  >{\raggedright\arraybackslash}p{(\columnwidth - 8\tabcolsep) * \real{0.2086}}
  >{\raggedright\arraybackslash}p{(\columnwidth - 8\tabcolsep) * \real{0.1519}}
  >{\raggedright\arraybackslash}p{(\columnwidth - 8\tabcolsep) * \real{0.2828}}
  >{\raggedright\arraybackslash}p{(\columnwidth - 8\tabcolsep) * \real{0.2526}}
  >{\raggedright\arraybackslash}p{(\columnwidth - 8\tabcolsep) * \real{0.1042}}@{}}
\toprule()
\begin{minipage}[b]{\linewidth}\raggedright
\textbf{Event Category}
\end{minipage} & \begin{minipage}[b]{\linewidth}\raggedright
\textbf{Period}
\end{minipage} & \begin{minipage}[b]{\linewidth}\raggedright
\textbf{Mean Swim-Out Time}
\end{minipage} & \begin{minipage}[b]{\linewidth}\raggedright
\textbf{SD Swim-Out Time}
\end{minipage} & \begin{minipage}[b]{\linewidth}\raggedright
\textbf{Count}
\end{minipage} \\
\midrule()
\endhead
\textbf{Short} & Covid & 1855 & 380 & 4694 \\
\textbf{Short} & Post-Covid & 1899 & 403 & 3726 \\
\textbf{Short} & Pre-Covid & 1696 & 415 & 37722 \\
\textbf{Long} & Covid & 4594 & 804 & 1130 \\
\textbf{Long} & Post-Covid & 4572 & 1035 & 1014 \\
\textbf{Long} & Pre-Covid & 4553 & 773 & 10201 \\
\textbf{Middle} & Covid & 2311 & 380 & 4658 \\
\textbf{Middle} & Post-Covid & 2315 & 394 & 3527 \\
\textbf{Middle} & Pre-Covid & 2216 & 436 & 26711 \\
\textbf{Sprint} & Covid & 854 & 295 & 7985 \\
\textbf{Sprint} & Post-Covid & 906 & 296 & 6152 \\
\textbf{Sprint} & Pre-Covid & 774 & 258 & 60871 \\
\bottomrule()
\end{longtable}

\textbf{Table 4.} Balance check across periods

\begin{longtable}[]{@{}
  >{\raggedright\arraybackslash}p{(\columnwidth - 12\tabcolsep) * \real{0.1395}}
  >{\raggedright\arraybackslash}p{(\columnwidth - 12\tabcolsep) * \real{0.1453}}
  >{\raggedright\arraybackslash}p{(\columnwidth - 12\tabcolsep) * \real{0.2543}}
  >{\raggedright\arraybackslash}p{(\columnwidth - 12\tabcolsep) * \real{0.0873}}
  >{\raggedright\arraybackslash}p{(\columnwidth - 12\tabcolsep) * \real{0.1311}}
  >{\raggedright\arraybackslash}p{(\columnwidth - 12\tabcolsep) * \real{0.1427}}
  >{\raggedright\arraybackslash}p{(\columnwidth - 12\tabcolsep) * \real{0.0999}}@{}}
\toprule()
\begin{minipage}[b]{\linewidth}\raggedright
\textbf{Period}
\end{minipage} & \begin{minipage}[b]{\linewidth}\raggedright
\textbf{Mean Total}
\end{minipage} & \begin{minipage}[b]{\linewidth}\raggedright
\textbf{Mean Swimming Time}
\end{minipage} & \begin{minipage}[b]{\linewidth}\raggedright
\textbf{Mean\\
Rank}\strut
\end{minipage} & \begin{minipage}[b]{\linewidth}\raggedright
\textbf{Mean Age}
\end{minipage} & \begin{minipage}[b]{\linewidth}\raggedright
\textbf{Mean Male}
\end{minipage} & \begin{minipage}[b]{\linewidth}\raggedright
\textbf{Count}
\end{minipage} \\
\midrule()
\endhead
\textbf{Covid} & 12091 & 1705 & 121 & 40.2 & 0.768 & 18467 \\
\textbf{Post-Covid} & 12573 & 1765 & 127 & 40.8 & 0.757 & 14419 \\
\textbf{Pre-Covid} & 11665 & 1599 & 165 & 38.1 & 0.796 & 135505 \\
\bottomrule()
\end{longtable}

\section{Theoretical Framework and Identification Strategy}

\emph{Effort and Positioning}

The swim segment is modelled as a two-stage game. In the first stage,
athletes choose swimming effort. In the second stage, they select a
drafting position within an inferred group, trading off the cost of
positioning against the benefit of reduced hydrodynamic drag. For
expositional simplicity, effort costs are always specified as linear in
effort, but this should be understood as a tractable monotonic
approximation rather than a literal description of physiology. If effort
costs were convex, the qualitative implications would remain similar,
although the equilibrium would place relatively more weight on interior
effort choices and could strengthen the value of efficient positioning.
The linear specification therefore simplifies the algebra without
changing the core strategic mechanism.

Drafting benefits are specified as a monotone function of proximity to
the leading swimmer, with diminishing marginal returns beyond a
positional threshold, consistent with prior empirical specifications
calibrated on the same underlying dataset (Reichel, 2025). The parameter
governing the accumulation rate of drafting gains is fixed at a rate of
decay of 0.5 in the baseline specification and should be interpreted as
a normalization rather than an estimated structural coefficient. This
choice is motivated by the lack of large-scale experimental
identification in comparable race settings and by hydrodynamic evidence
suggesting that drafting externalities decline nonlinearly with
distance, making a parsimonious convex decay specification appropriate
for the baseline model. This fixed value is not separately estimated; a
full sensitivity analysis varying $\lambda$ over a plausible range (e.g.,
0.3--0.7) is left for future work, and readers should treat the precise
curvature of the theory-based specification (Tables 10--12) as
illustrative for future and related research rather than fully
structurally estimated.

\emph{Main Identification Strategy}

The main identification challenge is that drafting position is
endogenous, arising from strategic sorting (e.g at the start),
endogenous group composition and non-random grouping by organizers (see sources below). To address these issues with a leave-one-out instrumental-variable strategy based on peer swimming ability. The logic is that peers' average ability
predicts the athlete's relative position within the inferred cluster,
but the leave-one-out construction removes the athlete's own
contribution and limits mechanical correlation with the outcome.

This strategy also helps to reduce bias from misclassified drafting
groups. If group assignment contains classical measurement error, the
estimated effect of drafting position would tend to be biased toward
zero. By instrumenting observed position with leave-one-out peer
ability, to recover variation in position that is tied to the local race
environment rather than to noisy classification alone. This does not
eliminate all measurement concerns, but it strengthens the causal
interpretation relative to a purely observational comparison.

\textbf{Table 5}. OLS estimation results - dependent variable: swim out
times

\begin{longtable}[]{@{}
  >{\raggedright\arraybackslash}p{(\columnwidth - 10\tabcolsep) * \real{0.3243}}
  >{\raggedright\arraybackslash}p{(\columnwidth - 10\tabcolsep) * \real{0.1613}}
  >{\raggedright\arraybackslash}p{(\columnwidth - 10\tabcolsep) * \real{0.0899}}
  >{\raggedright\arraybackslash}p{(\columnwidth - 10\tabcolsep) * \real{0.0899}}
  >{\raggedright\arraybackslash}p{(\columnwidth - 10\tabcolsep) * \real{0.1443}}
  >{\raggedright\arraybackslash}p{(\columnwidth - 10\tabcolsep) * \real{0.1903}}@{}}
\toprule()
\begin{minipage}[b]{\linewidth}\raggedright
\end{minipage} & \begin{minipage}[b]{\linewidth}\raggedright
\textbf{Estimate}
\end{minipage} &
\multicolumn{2}{>{\raggedright\arraybackslash}p{(\columnwidth - 10\tabcolsep) * \real{0.1797} + 2\tabcolsep}}{%
\begin{minipage}[b]{\linewidth}\raggedright
\textbf{Std. Error}
\end{minipage}} & \begin{minipage}[b]{\linewidth}\raggedright
\textbf{t-value}
\end{minipage} & \begin{minipage}[b]{\linewidth}\raggedright
\(Pr(\) \textbf{\textless t} \()\)
\end{minipage} \\
\midrule()
\endhead
\textbf{Drafter} & -23.8886 &
\multicolumn{2}{>{\raggedright\arraybackslash}p{(\columnwidth - 10\tabcolsep) * \real{0.1797} + 2\tabcolsep}}{%
4.1184} & -5.8004 & 0.0000 *** \\
\textbf{Pre-period $\times$ Drafter} & -8.5779 &
\multicolumn{2}{>{\raggedright\arraybackslash}p{(\columnwidth - 10\tabcolsep) * \real{0.1797} + 2\tabcolsep}}{%
4.6013} & -1.8642 & 0.0623 .\\
\textbf{Covid-year $\times$ Drafter} & 19.4150 &
\multicolumn{2}{>{\raggedright\arraybackslash}p{(\columnwidth - 10\tabcolsep) * \real{0.1797} + 2\tabcolsep}}{%
8.1469} & 2.3831 & 0.0172 * \\
\multicolumn{3}{@{}>{\raggedright\arraybackslash}p{(\columnwidth - 10\tabcolsep) * \real{0.5755} + 4\tabcolsep}}{%
\textbf{Observations}} &
\multicolumn{3}{>{\raggedright\arraybackslash}p{(\columnwidth - 10\tabcolsep) * \real{0.4245} + 4\tabcolsep}@{}}{%
168,391} \\
\multicolumn{3}{@{}>{\raggedright\arraybackslash}p{(\columnwidth - 10\tabcolsep) * \real{0.5755} + 4\tabcolsep}}{%
\textbf{Fixed-effects}} &
\multicolumn{3}{>{\raggedright\arraybackslash}p{(\columnwidth - 10\tabcolsep) * \real{0.4245} + 4\tabcolsep}@{}}{%
\begin{minipage}[t]{\linewidth}\raggedright
Athlete ID: 29,194,\\
Event ID: 1,339,

Cluster (Swim Group): 224\strut
\end{minipage}} \\
\multicolumn{3}{@{}>{\raggedright\arraybackslash}p{(\columnwidth - 10\tabcolsep) * \real{0.5755} + 4\tabcolsep}}{%
\textbf{Standard-errors}} &
\multicolumn{3}{>{\raggedright\arraybackslash}p{(\columnwidth - 10\tabcolsep) * \real{0.4245} + 4\tabcolsep}@{}}{%
Heteroskedasticity-robust} \\
\multicolumn{3}{@{}>{\raggedright\arraybackslash}p{(\columnwidth - 10\tabcolsep) * \real{0.5755} + 4\tabcolsep}}{%
\textbf{RMSE}} &
\multicolumn{3}{>{\raggedright\arraybackslash}p{(\columnwidth - 10\tabcolsep) * \real{0.4245} + 4\tabcolsep}@{}}{%
175.1} \\
\multicolumn{3}{@{}>{\raggedright\arraybackslash}p{(\columnwidth - 10\tabcolsep) * \real{0.5755} + 4\tabcolsep}}{%
\textbf{Adj.} \(\mathbf{R}^{\mathbf{2}}\)} &
\multicolumn{3}{>{\raggedright\arraybackslash}p{(\columnwidth - 10\tabcolsep) * \real{0.4245} + 4\tabcolsep}@{}}{%
0.9683} \\
\multicolumn{3}{@{}>{\raggedright\arraybackslash}p{(\columnwidth - 10\tabcolsep) * \real{0.5755} + 4\tabcolsep}}{%
\textbf{Within} \(\mathbf{R}^{\mathbf{2}}\)} &
\multicolumn{3}{>{\raggedright\arraybackslash}p{(\columnwidth - 10\tabcolsep) * \real{0.4245} + 4\tabcolsep}@{}}{%
0.0035} \\
\bottomrule()
\end{longtable}

\textbf{Notes}: p~\textless{} 0.001 ***,~p~\textless{} 0.01
**,~p~\textless{} 0.05 *,~p~\textless{} 0.1 .

\emph{Exclusion Restriction}

The key exclusion restriction is that peer ability affects an athlete's
final race outcome only through drafting position, not through other
direct channels (Angrist \& Pischke, 2009; Wooldridge, 2010). A
potential threat is psychological spillover: being surrounded by faster
peers could motivate athletes or influence pacing independently of hydrodynamic effects. This concern is plausible, but the dominant channel in open-water swimming is
likely physical rather than motivational, given the short, crowded and
contact-constrained nature of the segment (Chatard et al., 2003;
Delextrat et al., 2003). Second, the empirical design can partly
separate these mechanisms by focusing on within-cluster positional
variation and by exploiting the COVID-era drafting bans, which altered
the feasibility of drafting more directly than peer composition itself
(Reichel, 2025) (Table 5).

\emph{Structural Potential Outcomes Framework}

Let \(Y_{ie}(d)\) denote the potential outcome for athlete \(i\) in
event \(e\) had they assumed drafting position d. Then:

\[Y_{ie\ }: = {\ Y}_{ie}\left( D_{ieg} \right)\]

The causal estimand compares each drafting position to the leader is:

\[ATE(d)\ : = \mathbb{E}\left\lbrack Y_{\mathfrak{i}e}(d) - Y_{\mathfrak{i}e}(1) \right\rbrack\ \text{~for~}d < Threshold\]

But due to exogenous variation, sorting, selection into drafting
positions, required unconfoundedness fails:

\[\mathbb{E}\left\lbrack Y_{ie} \mid D_{ieg} = d \right\rbrack \neq \mathbb{E}\left\lbrack Y_{ie}(\text{\ }d) \right\rbrack\]

To correct for this via an IV strategy with a theoretically derived
concave transformation of drafting depth.

\emph{Endogeneity and identification strategy}

Endogeneity in drafting position arises from:

\begin{enumerate}
\def\labelenumi{\roman{enumi}.}
\item
  Strategic Sorting: Higher-ability athletes self-select into favorable
  positions.
\item
  Endogenous Grouping: Group composition reflects latent ability.
\item
  Non-random Grouping: Organizers may group by prior race performance
  (e.g. start waves) or likelihood of receiving a time penalty during a
  race.
\end{enumerate}

Hence:

\[Cov\left( D_{ieg},\varepsilon_{ie} \right) \neq 0\]

\emph{Leave-One-Out Instrumental variable}

Let the instrument be defined as

\[Z_{ie\ }: = \frac{1}{\left| \mathcal{G}_{ie}^{- i} \right|}\sum_{j \in \mathcal{G}_{ie}^{- i}}^{}\mspace{2mu}{\text{\ }S}_{je}\]

\hfill\break
This captures the swim ability of groups in athlete i\textquotesingle s
drafting within the current group, excluding i. This leave-one-out
(LOO/LOTO) instrument captures the average group ability while avoiding
mechanical reflection bias. By excluding athlete, one removes the
mechanical correlation between the instrument and the outcome. This
strategy aligns with methods used in social interaction and group effect
models to address endogeneity, e.g., Bramoulle (2009); Angrist (2014).
Related applications include Leave-One-Out instruments in production
function estimation.

\emph{Drafting benefit transformation from theoretic model}

Instead of using raw position, a structural regressor from theory is
constructed as:

\[B_{ieg} = \left\{ \begin{matrix}
0 & \text{~if~}D_{ieg} \leq 3 \\
\gamma\left( 1 - e^{- \lambda\left( D_{ieg} - 3 \right)} \right) & \text{~if~}D_{ieg} > 3 \\
\end{matrix} \right.\ \]

\(N\)ormalized using fixed parameters of
\(\gamma = 1,\ \ \lambda = 0.5\) and set as described above and for the
sake of simplicity and brevity. This transformation captures the
increasing-but-saturating drafting benefits posited by the theoretical
model.

\emph{\hfill\break
Two-Stage Least Squares (2SLS) Group-IV Estimation}

To identify the causal effect of within-group drafting benefit on race
performance, a two-stage least squares (2SLS) model using a
leave-one-out group ability instrument is estimated. This setup accounts
for endogeneity in observed drafting positions, which are determined by
both athlete ability and strategic group formation. Let index athletes,
index events, and index drafting clusters within an event. The key
endogenous variable is, a nonlinear transformation of the
athlete\textquotesingle s drafting position, derived from the structural
theory. To isolate exogenous variation in this drafting benefit using
the average swimming ability of other athletes in the same group as an
instrument.

First Stage Equation: To instrument for Drafting Position the drafting
position as a function of group swim ability and controls estimating
equation is specified as:

\[D_{ieg} = \pi_{0} + \pi_{1}Z_{ie} + \pi_{2}S_{ie} + X_{ie}^{'}\pi_{3} + \lambda_{e} + \alpha_{i} + \alpha_{g} + u_{ieg}
\]

\begin{enumerate}
\def\labelenumi{\roman{enumi}.}
\item
  \(D_{\text{ieg~}}\) : observed drafting position of athlete \(i\) in
  event \(e\), group \(g\);
\item
  \(Z_{ie}\mathfrak{i}\) : leave-one-out group average swim time
  (excluding athlete );
\item
  \(S_{\mathfrak{i}e}\) : own swim time of athlete \(\mathfrak{i}\),
  capturing own ability;
\item
  \(X_{ie}\) : vector of additional athlete- or event-level covariates
  (e.g., age, gender);
\item
  \(\lambda_{e}\) : event fixed effects, capturing course
  characteristics and environmental conditions;
\item
  \(\alpha_{i}\) : athlete fixed effects, absorbing time-invariant
  ability heterogeneity;
\item
  \(\alpha_{g}\) : group fixed effects, accounting for shared
  cluster-level dynamics;
\item
  \(u_{\text{ieg~}}\) : idiosyncratic first-stage error term.
\end{enumerate}

Second Stage: Estimating the Effect of Drafting Benefit on Performance
Using

\[{\widehat{B}}_{\text{ieg~}} = f\left( {\widehat{D}}_{\text{ieg~}} \right)\]

the fitted drafting benefit to validate the theoretical model or using
directly, the following log-outcome transformed equation is then
estimated:

\[log\left( \text{rank~}_{ie} + 1 \right) = \beta_{1}{\widehat{B}}_{ieg} + \beta_{2} \cdot \text{~leader~}_{ieg} + X_{ie}^{'}\beta_{3} + \lambda_{e} + \alpha_{i} + \alpha_{g} + \varepsilon_{ieg}\]

Where:

\begin{enumerate}
\def\labelenumi{\roman{enumi}.}
\item
  \(log\left( rank \right.\ \left. \ \ _{ie} + 1 \right)\) :
  log-transformed within-event finishing rank of athlete
  \(\mathfrak{i}\), adjusted to avoid log-zero;
\item
  \({\widehat{B}}_{ieg}\) : instrumented drafting benefit, based on the
  theoretical transformation of \({\widehat{D}}_{ieg}\);
\item
  leader \(\ _{ieg}\) : indicator variable for being in the frontmost
  drafting position within group \(g\);
\item
  \(X_{ie},\lambda_{e},\alpha_{i},\alpha_{g}\) : as defined above;
\item
  \(\varepsilon_{\text{ieg~}}\) : second-stage error term.
\end{enumerate}

\emph{\hfill\break
Identification Strategy And Exogeneity}

This 2SLS framework identifies the local average treatment effect (LATE)
of drafting benefit for compliers-athletes whose position is affected by
group group composition.

Identification Strategy Identification Assumptions:

\begin{enumerate}
\def\labelenumi{\roman{enumi}.}
\item
  Instrument relevance:
  \(Cov\left( Z_{\text{ie~}},D_{\text{ieg~}} \right) \neq 0\);
\item
  Instrument exogeneity:
  \(\mathbb{E}\left\lbrack \varepsilon_{ieg} \mid Z_{ie},{\text{\ }S}_{ie},X_{ie},\lambda_{e},\alpha_{i},\alpha_{g} \right\rbrack = 0\);
\item
  Monotonicity: An increase in group ability (faster groups) weakly
  shifts athletes to deeper drafting positions.
\end{enumerate}

This empirical structure provides consistent estimates of the causal
effect of drafting benefit on race outcomes while accounting for
sorting, selection, and unobserved heterogeneity through fixed effects
and instrument design.

\emph{Structural Validation and Interpretation}

The coefficient \(\beta_{1}\) reflects the average marginal return to
drafting benefit. The increasing and concave structure implies that:

\begin{enumerate}
\def\labelenumi{\roman{enumi}.}
\item
  Greatest positive net gains occur when transitioning from front
  positions backwards;
\item
  Diminishing returns occur deeper into the pack;
\item
  The optimal position is interior, as predicted by the theoretical
  model.
\end{enumerate}

Empirical evidence of this shape supports the game\textquotesingle s
interior equilibrium logic and nonlinear (dis-)utility (drafting benefit
structure) topology.

\emph{Potential Extension: Projected Panel IV}

For robustness and within-athlete variation, One can define a projected
instrument using historical group characteristics from the largest group
containing athlete , e.g.:

\[Z_{{ie}^{'}}^{\text{Projected~}} = \mathbb{E}_{e}\left\lbrack Z_{\mathfrak{ie}} \mid \mathfrak{i} \in g_{\mathfrak{ie}},e \neq e^{'} \right\rbrack\]

This makes more use of our rich panel data set with repeated athletes
across multiple events to increase within-athlete variation.\\
\emph{\hfill\break
Truncation Bias}

Restricting the analysis to the capped outcome specification, in which
the dependent variable is transformed, helps focus on performance among
finishers while reducing the influence of extreme outliers. However,
this specification is best interpreted as a supplementary check rather
than the main basis for inference, because truncation may introduce
selection concerns.

\section{Results}

\emph{Non-Linear Concave Drafting Position Effects}

\textbf{Table 6.} Instrumental variables estimates of drafting effects
on performance

\begin{longtable}[]{@{}
  >{\raggedright\arraybackslash}p{(\columnwidth - 6\tabcolsep) * \real{0.2803}}
  >{\raggedright\arraybackslash}p{(\columnwidth - 6\tabcolsep) * \real{0.2254}}
  >{\raggedright\arraybackslash}p{(\columnwidth - 6\tabcolsep) * \real{0.2471}}
  >{\raggedright\arraybackslash}p{(\columnwidth - 6\tabcolsep) * \real{0.2471}}@{}}
\toprule()
\begin{minipage}[b]{\linewidth}\raggedright
\end{minipage} & \begin{minipage}[b]{\linewidth}\raggedright
\textbf{(1) Full Sample}
\end{minipage} & \begin{minipage}[b]{\linewidth}\raggedright
\textbf{(2) Group Size \textless{} 15}
\end{minipage} & \begin{minipage}[b]{\linewidth}\raggedright
\textbf{(3) Group Size \textless{} 10}
\end{minipage} \\
\midrule()
\endhead
\textbf{Drafting Position (Fit)} & -0.039*** (0.007) & -1.729*** (0.491)
& -2.496** (0.910) \\
\textbf{Leader} & -0.301*** (0.052) & -5.328*** (1.526) & -6.159**
(2.235) \\
\textbf{Observations} & 45,215 & 26,357 & 21,190 \\
\textbf{Athletes (FE)} & 15,762 & 11,919 & 10,534 \\
\textbf{Events (FE)} & 779 & 793 & 786 \\
\textbf{Clusters (FE)} & 149 & 149 & 149 \\
\textbf{RMSE} & 1.036 & 3.601 & 3.339 \\
\textbf{Adj.} \(\mathbf{R}^{\mathbf{2}}\) & -0.128 & -11.8 & -13.6 \\
\textbf{Within} \(\mathbf{R}^{\mathbf{2}}\) & -1.979 & -33.1 & -46.8 \\
\textbf{First Stage F-stat} & 998.4 & 24.7 & 14.6 \\
\textbf{Wu-Hausman} \(\mathbf{p}\)\textbf{-value} & 0.0000 & 0.0000 &
0.0000 \\
\bottomrule()
\end{longtable}

\textbf{Notes}: p~\textless{} 0.001 ***,~p~\textless{} 0.01
**,~p~\textless{} 0.05 *. Estimates are based on two-stage least squares
(2SLS), with drafter position treated as endogenous and instrumented by
the leave-one-out cluster-level mean swimming ability. The dependent
variable is , where denotes the centered within-event finishing rank of
athlete . All models include fixed effects for athlete, event, and
drafting cluster. Standard errors are clustered at the event level. In
Column (1), the coefficient on drafter position is small but
significantly negative, suggesting that being farther behind in a
drafting cluster is associated with better performance (lower log rank).
The effect increases substantially in magnitude and significance in
Columns (2) and (3), which restrict the sample to smaller clusters,
implying stronger drafting dynamics in such settings. The coefficient on
the Leader dummy is also consistently negative and statistically
significant, indicating that athletes who lead clusters tend to finish
with better ranks. This likely reflects a combination of superior
ability and favorable positioning strategy. Some observations were
excluded due to missing or invalid values of the dependent variable,
including cases where was undefined or produced NaNs. This may stem from
negative centering; future versions could offset for robustness.

\textbf{Table 7.} Instrumental variables estimates of interpretable
drafting effects on performance

\begin{longtable}[]{@{}
  >{\raggedright\arraybackslash}p{(\columnwidth - 6\tabcolsep) * \real{0.2841}}
  >{\raggedright\arraybackslash}p{(\columnwidth - 6\tabcolsep) * \real{0.2192}}
  >{\raggedright\arraybackslash}p{(\columnwidth - 6\tabcolsep) * \real{0.2551}}
  >{\raggedright\arraybackslash}p{(\columnwidth - 6\tabcolsep) * \real{0.2416}}@{}}
\toprule()
\begin{minipage}[b]{\linewidth}\raggedright
\end{minipage} & \begin{minipage}[b]{\linewidth}\raggedright
\textbf{(1) Full Sample}
\end{minipage} & \begin{minipage}[b]{\linewidth}\raggedright
\textbf{(2) Group Size \textless{} 10}
\end{minipage} & \begin{minipage}[b]{\linewidth}\raggedright
\textbf{(3) Group Size \textless{} 5}
\end{minipage} \\
\midrule()
\endhead
\textbf{Drafter Position (Fit)} & -0.011* (0.0053) & -0.972*** (0.185) &
-2.378* (1.073) \\
\textbf{Observations} & 34,061 & 16,035 & 9,316 \\
\textbf{Athletes (FE)} & 12,181 & 8,408 & 5,856 \\
\textbf{Events (FE)} & 438 & 523 & 531 \\
\textbf{Clusters (FE)} & 178 & 178 & 176 \\
\textbf{RMSE} & 0.771 & 1.161 & 1.176 \\
\textbf{Adj.} \(\mathbf{R}^{\mathbf{2}}\) & 0.641 & -0.166 & -0.674 \\
\textbf{Within} \(\mathbf{R}^{\mathbf{2}}\) & -0.905 & -5.925 &
-8.453 \\
\textbf{First Stage F-stat} & 1,115.6 & 51.4 & 12.7 \\
\textbf{First Stage} \(\mathbf{p}\)\textbf{-value} & 0.0000 & 0.0000 &
0.0000 \\
\textbf{Wu-Hausman p-value} & 0.0000 & 0.0000 & 0.0000 \\
\bottomrule()
\end{longtable}

\textbf{Notes}: p~\textless{} 0.001 ***,~p~\textless{} 0.01
**,~p~\textless{} 0.05 *. Two-stage least squares (2SLS) estimates of Drafter Position (Fit),
where is the centered finishing rank of athlete within event. The
endogenous regressor is the athlete\textquotesingle s drafter position ,
instrumented using the cluster-level Group mean swim performance
(leave-one-out). No additional covariates are included. Fixed effects
for athlete, cluster, and event are used. Standard errors are clustered
at the event level. In Column (1), the coefficient is small but
statistically significant, indicating modest benefits from trailing
farther behind. The magnitude and significance of the effect increase in
Columns (2) and (3), which restrict to smaller drafting clusters. In
these subsamples, being positioned farther back in the pack is
associated with significantly better performance, consistent with
drafting theory. The precision of the estimates improves in Column (2)
due to stronger first-stage relevance compared to Column (3). All models
pass the Wu-Hausman test, supporting the endogeneity of drafter
position. Some observations were dropped in Columns (2) and (3), respectively, due to missing or invalid values of the
dependent variable, including cases where is undefined (e.g., due to
negative centering).

Table 7 shows that the coefficient on drafting position is negative in
the full sample and remains negative in smaller groups. Because the
outcome is as, a negative coefficient implies a lower finishing rank,
which means better performance rather than a penalty. The magnitude of
the estimated effects is strongest in smaller groups, especially in very
small clusters. However, the very large percentage changes implied by the capped
specification should be interpreted cautiously. In groups with fewer
than five athletes, the implied improvement is economically large and
may reflect the combination of a small effective sample, greater sensitivity to
specification choice, and possible weak-instrument bias rather than a
pure structural hydrodynamic effect. OLS estimates using position-dummy
fixed-effects models are provided in Tables 8--9.

\textbf{Table 8.} OLS regression results considering all participants in
non-leading swim-group roles as potential drafting beneficiaries

\begin{longtable}[]{@{}
  >{\raggedright\arraybackslash}p{(\columnwidth - 8\tabcolsep) * \real{0.1734}}
  >{\raggedright\arraybackslash}p{(\columnwidth - 8\tabcolsep) * \real{0.2083}}
  >{\raggedright\arraybackslash}p{(\columnwidth - 8\tabcolsep) * \real{0.2197}}
  >{\raggedright\arraybackslash}p{(\columnwidth - 8\tabcolsep) * \real{0.2070}}
  >{\raggedright\arraybackslash}p{(\columnwidth - 8\tabcolsep) * \real{0.1917}}@{}}
\toprule()
\begin{minipage}[b]{\linewidth}\raggedright
\end{minipage} & \begin{minipage}[b]{\linewidth}\raggedright
\textbf{(1) Pre-Covid Sample (} \(\leq 2020\) \textbf{)}
\end{minipage} & \begin{minipage}[b]{\linewidth}\raggedright
\textbf{(2) Covid Sample (2020, 2021, 2022)}
\end{minipage} & \begin{minipage}[b]{\linewidth}\raggedright
\textbf{(3) Post-Covid Sample (} \(\geq\) \textbf{2023)}
\end{minipage} & \begin{minipage}[b]{\linewidth}\raggedright
\textbf{(4) Full Sample (2010-2024)}
\end{minipage} \\
\midrule()
\endhead
\multicolumn{5}{@{}>{\raggedright\arraybackslash}p{(\columnwidth - 8\tabcolsep) * \real{1.0000} + 8\tabcolsep}@{}}{%
\textbf{Panel A: OLS Estimations (Swim Group Def.: Max 5 sec apart,
Cluster (Swim Group) Complete Linkage, Subsample based on Athletes:
Competing in all 3 Periods)}} \\
\textbf{Observations} & 29,750 & 9,796 & 6,941 & 46,487 \\
\textbf{Fixed Effects} & Event, Athlete & Event, Athlete & Event,
Athlete & Event, Athlete \\
\textbf{Clustered SE} & Event ID & Event ID & Event ID & Event ID \\
\textbf{Drafter} & -19.476*** & -13.061*** & -15.648** & -18.569*** \\
\textbf{Standard Error} & (2.413) & (3.517) & (4.668) & (1.855) \\
\textbf{RMSE} & 171.3 & 147.9 & 129.7 & 178.9 \\
\textbf{Adjusted} \(\mathbf{R}^{\mathbf{2}}\) & 0.9709 & 0.9742 & 0.9793
& 0.9698 \\
\textbf{Within} \(\mathbf{R}^{\mathbf{2}}\) & 0.0021 & 0.0012 & 0.0020 &
0.0019 \\
\multicolumn{5}{@{}>{\raggedright\arraybackslash}p{(\columnwidth - 8\tabcolsep) * \real{1.0000} + 8\tabcolsep}@{}}{%
\textbf{Group) Complete Linkage)}} \\
\textbf{Observations} & 135,505 & 18,467 & 14,419 & 168,391 \\
\textbf{Fixed Effects} & Event, Athlete & Event, Athlete & Event,
Athlete & Event, Athlete \\
\textbf{Clustered SE} & Event ID & Event ID & Event ID & Event ID \\
\textbf{Drafter} & -22.892*** & -18.307*** & -22.388*** & -23.047*** \\
\textbf{Standard Error} & (1.459) & (4.179) & (4.365) & (1.330) \\
\textbf{RMSE} & 166.6 & 140.1 & 116.1 & 179.4 \\
\textbf{Adjusted} \(\mathbf{R}^{\mathbf{2}}\) & 0.9717 & 0.9707 & 0.9781
& 0.9668 \\
\textbf{Within} \(\mathbf{R}^{\mathbf{2}}\) & 0.0035 & 0.0024 & 0.0045 &
0.0030 \\
\multicolumn{5}{@{}>{\raggedright\arraybackslash}p{(\columnwidth - 8\tabcolsep) * \real{1.0000} + 8\tabcolsep}@{}}{%
\textbf{Panel C: OLS Estimations (Swim Group Def.: Max 15 sec apart,
Cluster (Swim Group) Complete Linkage, Subsample based on Athletes: At
Least} \(\mathbf{3}\) \textbf{Swimmers per Group)}} \\
\textbf{Observations} & 111,880 & 14,262 & 11,011 & 137,153 \\
\textbf{Fixed Effects} & Event, Athlete & Event, Athlete & Event,
Athlete & Event, Athlete \\
\textbf{Clustered SE} & Event ID & Event ID & Event ID & Event ID \\
\textbf{Drafter} & -5.348*** & -3.206 & -9.764* & -5.339*** \\
\textbf{Standard Error} & (0.951) & (2.620) & (4.082) & (0.841) \\
\textbf{RMSE} & 122.7 & 92.1 & 76.5 & 123.4 \\
\textbf{Adjusted} \(\mathbf{R}^{\mathbf{2}}\) & 0.9809 & 0.9826 & 0.9860
& 0.9803 \\
\textbf{Within} \(\mathbf{R}^{\mathbf{2}}\) & 0.00024 & 0.00013 &
0.00139 & 0.00024 \\
\multicolumn{5}{@{}>{\raggedright\arraybackslash}p{(\columnwidth - 8\tabcolsep) * \real{1.0000} + 8\tabcolsep}@{}}{%
\textbf{Note: ***}
\(\mathbf{p} < \mathbf{0}.\mathbf{001},\ ^{**}\mathbf{p} < \mathbf{0}.\mathbf{01},\ ^{*}\mathbf{p} < \mathbf{0}.\mathbf{05}\)} \\
\bottomrule()
\end{longtable}

\begin{longtable}[]{@{}
  >{\raggedright\arraybackslash}p{(\columnwidth - 8\tabcolsep) * \real{0.3251}}
  >{\raggedright\arraybackslash}p{(\columnwidth - 8\tabcolsep) * \real{0.2018}}
  >{\raggedright\arraybackslash}p{(\columnwidth - 8\tabcolsep) * \real{0.1620}}
  >{\raggedright\arraybackslash}p{(\columnwidth - 8\tabcolsep) * \real{0.1505}}
  >{\raggedright\arraybackslash}p{(\columnwidth - 8\tabcolsep) * \real{0.1607}}@{}}
\toprule()
\multicolumn{5}{@{}>{\raggedright\arraybackslash}p{(\columnwidth - 8\tabcolsep) * \real{1.0000} + 8\tabcolsep}@{}}{%
\begin{minipage}[b]{\linewidth}\raggedright
\textbf{Panel D: OLS Estimation (Dependent Variable: Swim-Out Time,
Clustered SE at Athlete Level, Swim Group Def.: Max 5 sec . apart,
Cluster (Swim Group) Single Linkage)}
\end{minipage}} \\
\midrule()
\endhead
\textbf{Fixed Effects}

\textbf{Clustered SE} &
\multicolumn{4}{>{\raggedright\arraybackslash}p{(\columnwidth - 8\tabcolsep) * \real{0.6749} + 6\tabcolsep}@{}}{%
168,391

Athlete \((29,194)\), Event \((1,339)\)

Athlete ID} \\
\textbf{Variable} & Estimate & Std. Error & t value & \(Pr( < t)\) \\
\textbf{Pre-period $\times$ Drafter} & -51.680*** & 1.841 & -28.065 &
0.00000 \\
& & & & \\
\textbf{Post-period $\times$ Drafter} & 15.138*** & 3.629 & 4.171 & 0.00000 \\
\multicolumn{5}{@{}>{\raggedright\arraybackslash}p{(\columnwidth - 8\tabcolsep) * \real{1.0000} + 8\tabcolsep}@{}}{%
\textbf{RMSE}

\textbf{178.8}} \\
\textbf{Adjusted} \(\mathbf{R}^{\mathbf{2}}\) & & & 0.9670 & \\
\textbf{Within} \(\mathbf{R}^{\mathbf{2}}\) & &
\multicolumn{3}{>{\raggedright\arraybackslash}p{(\columnwidth - 8\tabcolsep) * \real{0.4732} + 4\tabcolsep}@{}}{%
0.0099} \\
\multicolumn{5}{@{}>{\raggedright\arraybackslash}p{(\columnwidth - 8\tabcolsep) * \real{1.0000} + 8\tabcolsep}@{}}{%
\textbf{Panel E: OLS Estimation (Dependent Variable: Swim-Out Time,
Additional Controls, Swim Group Def.: Max 5 sec . apart, Cluster (Swim
Group) Single Linkage)}} \\
\textbf{Observations}

\textbf{Fixed Effects}

\textbf{Clustered SE} &
\multicolumn{4}{>{\raggedright\arraybackslash}p{(\columnwidth - 8\tabcolsep) * \real{0.6749} + 6\tabcolsep}@{}}{%
168,391

Athlete \((29,194)\), Event \((1,339)\)

Athlete ID} \\
\textbf{Pre-period $\times$ Drafter} & -29.858*** & 1.880 & -15.886 &
0.00000 \\
\textbf{Post-period $\times$ Drafter} & 14.012*** & 3.627 & 3.864 & 0.00011 \\
\textbf{Drafter Position} & -0.173*** & 0.018 & -9.398 & 0.00000 \\
\textbf{Cluster (Swim Group)} & 1.669*** & 0.107 & 15.564 & 0.00000 \\
\multicolumn{5}{@{}>{\raggedright\arraybackslash}p{(\columnwidth - 8\tabcolsep) * \real{1.0000} + 8\tabcolsep}@{}}{%
} \\
\multicolumn{5}{@{}>{\raggedright\arraybackslash}p{(\columnwidth - 8\tabcolsep) * \real{1.0000} + 8\tabcolsep}@{}}{%
\textbf{Panel F: OLS Estimation (Dependent Variable: Swim-Out Time,
Clustered SE at Athlete \& Event Level, Swim Group Def.: Max 5 sec .
apart, Cluster (Swim Group) Single Linkage)}} \\
\textbf{Observations}

\textbf{Fixed Effects}

\textbf{Clustered SE} &
\multicolumn{4}{>{\raggedright\arraybackslash}p{(\columnwidth - 8\tabcolsep) * \real{0.6749} + 6\tabcolsep}@{}}{%
168,391

Athlete \((29,194)\), Event \((1,339)\)

TWFE (Athlete ID, Event ID)} \\
\textbf{Pre-period $\times$ Drafter} & -29.967*** & 2.433 & -12.317 & 0.0000 \\
\textbf{Post-period $\times$ Drafter} & 12.543** & 4.357 & 2.879 & 0.0041 \\
\textbf{Drafter Position} & -0.271** & 0.091 & -2.978 & 0.0030 \\
\textbf{Cluster (Swim Group)} & 1.518*** & 0.124 & 12.285 & 0.00000 \\
\textbf{Race Rank} & 0.402*** & 0.030 & 13.316 & 0.00000 \\
& & & & \\
\multicolumn{5}{@{}>{\raggedright\arraybackslash}p{(\columnwidth - 8\tabcolsep) * \real{1.0000} + 8\tabcolsep}@{}}{%
\textbf{Note: ***}
\(\mathbf{p} < \mathbf{0}.\mathbf{001},\ ^{**}\mathbf{p} < \mathbf{0}.\mathbf{01},*\mathbf{p} < \mathbf{0}.\mathbf{05}\)} \\
\multicolumn{5}{@{}>{\raggedright\arraybackslash}p{(\columnwidth - 8\tabcolsep) * \real{1.0000} + 8\tabcolsep}@{}}{%
\textbf{Panel G: OLS Estimation (Additional Swim Group Cluster-Level
Fixed Effects, Robust SE, Swim Group Def.: Max 5 sec . apart, Cluster
(Swim Group) Single Linkage)}} \\
\textbf{Fixed Effects}

\textbf{Clustered SE} &
\multicolumn{4}{>{\raggedright\arraybackslash}p{(\columnwidth - 8\tabcolsep) * \real{0.6749} + 6\tabcolsep}@{}}{%
Cluster (Swim Group) \((224)\), Athlete \((29,194)\), Event \((1,339)\)
Robust SE (HC1)} \\
\textbf{Pre-period $\times$ Drafter} & -31.401*** & 1.652 & -19.013 & 0.0000 \\
\textbf{Post-period $\times$ Drafter} & 12.562*** & 3.594 & 3.495 & 0.00047 \\
\textbf{Drafter Position} & -0.180*** & 0.018 & -10.277 & 0.0000 \\
\textbf{RMSE} &
\multicolumn{4}{>{\raggedright\arraybackslash}p{(\columnwidth - 8\tabcolsep) * \real{0.6749} + 6\tabcolsep}@{}}{%
175.0} \\
\textbf{Adjusted} \(\mathbf{R}^{\mathbf{2}}\) & & & 0.9684 & \\
\textbf{Within} \(\mathbf{R}^{\mathbf{2}}\) & & & 0.0045 & \\
\bottomrule()
\end{longtable}

\textbf{Table 9}. OLS Estimation Results for Swim-Out Times Under
Different Model Specifications. \textbf{Notes}: p~\textless{} 0.001
***,~p~\textless{} 0.01 **,~p~\textless{} 0.05 *. This table presents
results from OLS regressions estimating the effects of drafting on
Swim-out times performance. (lower is better) Panel D estimates the
effects using an interaction terms specification, clustering at the
athlete level. Panel E introduces additional controls such as the
swimmer\textquotesingle s relative position in the drafting group
(drafter\_position) and cluster ID denoting the swim group. Panel F
further refines the estimation by clustering standard errors at both the
athlete and event level. The\\
dependent variable is the Swim-Out Time time in all models. Panel G
extends the analysis by including cluster-level fixed effects, which
account far within-group variations in drafting performance, while using
HC1 standard errors robust to heteroskedasticity.

\emph{Theory-Based Specification}

The theory-based specification in Tables 10--12 supports the same
qualitative pattern. The drafting-benefit term enters negatively and
significantly across specifications, which is consistent with improved
performance as drafting benefits increase. The effect is strongest in
smaller groups and remains negative in broader samples, suggesting that
drafting benefits are real but attenuate as cluster size grows and
positional gains saturate. The leader dummy is also negative and
statistically significant in several specifications. Given the IV
design, this result should be interpreted as evidence that front
positions are not necessarily costly in themselves and may instead
capture small tactical or congestion-related advantages. This
interpretation is more consistent with the estimated signs than a claim
that leading is uniformly disadvantageous.

\textbf{Table 10.} Instrumental variables estimates of theory-based
drafting effects on performance.

\begin{longtable}[]{@{}
  >{\raggedright\arraybackslash}p{(\columnwidth - 6\tabcolsep) * \real{0.2766}}
  >{\raggedright\arraybackslash}p{(\columnwidth - 6\tabcolsep) * \real{0.2267}}
  >{\raggedright\arraybackslash}p{(\columnwidth - 6\tabcolsep) * \real{0.2484}}
  >{\raggedright\arraybackslash}p{(\columnwidth - 6\tabcolsep) * \real{0.2484}}@{}}
\toprule()
\begin{minipage}[b]{\linewidth}\raggedright
\end{minipage} & \begin{minipage}[b]{\linewidth}\raggedright
\textbf{(1) All Groups}
\end{minipage} & \begin{minipage}[b]{\linewidth}\raggedright
\textbf{(2) Group Size \textless{} 20}
\end{minipage} & \begin{minipage}[b]{\linewidth}\raggedright
\textbf{(3) Group Size \textless{} 10}
\end{minipage} \\
\midrule()
\endhead
\textbf{Drafting Benefit (Fit)} & -6.650*** (0.654) & -10.662*** (1.871)
& -16.711** (5.239) \\
\textbf{Leader} & -2.845*** (0.288) & -3.676*** (0.659) & -3.980**
(1.256) \\
\textbf{Observations} & 45,215 & 29,073 & 21,190 \\
\textbf{Athletes (FE)} & 15,762 & 12,544 & 10,534 \\
\textbf{Events (FE)} & 779 & 789 & 786 \\
\textbf{Clusters (FE)} & 149 & 149 & 149 \\
\textbf{RMSE} & 1.717 & 2.570 & 3.165 \\
\textbf{Adj.} \(\mathbf{R}^{\mathbf{2}}\) & -2.100 & -6.635 & -12.100 \\
\textbf{Within} \(\mathbf{R}^{\mathbf{2}}\) & -7.186 & -22.200 &
-42.000 \\
\textbf{First Stage F-stat} & 254.7 & 53.0 & 16.3 \\
\textbf{First Stage p -value} & 0.000 & 0.000 & 0.000 \\
\textbf{Wu-Hausman p-value} & 0.000 & 0.000 & 0.000 \\
\bottomrule()
\end{longtable}

\textbf{Notes}: p~\textless{} 0.001 ***,~p~\textless{} 0.01
**,~p~\textless{} 0.05 *. Each column reports two-stage least squares
(2SLS) estimates of , where is the athlete\textquotesingle s centered
within-event rank. The endogenous variable is a theory-based drafting
benefit:

\[B\left( d_{i} \right) = \left\{ \begin{matrix}
0, & \text{~if~}d_{i} \leq 3 \\
\gamma\left( 1 - e^{- \lambda\left( d_{i} - 3 \right)} \right), & \text{~if~}d_{i} > 3 \\
\end{matrix} \right.\ 
\]

\textbf{Table 11.} 2SLS estimates using theory-implied drafting benefit

\begin{longtable}[]{@{}
  >{\raggedright\arraybackslash}p{(\columnwidth - 12\tabcolsep) * \real{0.1789}}
  >{\raggedright\arraybackslash}p{(\columnwidth - 12\tabcolsep) * \real{0.1261}}
  >{\raggedright\arraybackslash}p{(\columnwidth - 12\tabcolsep) * \real{0.1345}}
  >{\raggedright\arraybackslash}p{(\columnwidth - 12\tabcolsep) * \real{0.1261}}
  >{\raggedright\arraybackslash}p{(\columnwidth - 12\tabcolsep) * \real{0.1352}}
  >{\raggedright\arraybackslash}p{(\columnwidth - 12\tabcolsep) * \real{0.1640}}
  >{\raggedright\arraybackslash}p{(\columnwidth - 12\tabcolsep) * \real{0.1352}}@{}}
\toprule()
\begin{minipage}[b]{\linewidth}\raggedright
\end{minipage} & \begin{minipage}[b]{\linewidth}\raggedright
\textbf{(1) All Groups}
\end{minipage} & \begin{minipage}[b]{\linewidth}\raggedright
\textbf{(2) Group} \(\mathbf{>}\) \textbf{1, No Leader}
\end{minipage} & \begin{minipage}[b]{\linewidth}\raggedright
\textbf{(3) Group \textless{} 20}
\end{minipage} & \begin{minipage}[b]{\linewidth}\raggedright
\textbf{(4) Group \textless{} 20, No Leader}
\end{minipage} & \begin{minipage}[b]{\linewidth}\raggedright
\textbf{(5) Group \textless{} 10}
\end{minipage} & \begin{minipage}[b]{\linewidth}\raggedright
\textbf{(6) Group \textless{} 10, No Leader}
\end{minipage} \\
\midrule()
\endhead
\textbf{Theoretic Drafting Benefit Fit} \(B\left( d_{i} \right)\) &
-6.650*** (0.654) & -2.871*** (0.267) & -4.836*** (0.697) & -3.609***
(0.468) & -9.615*** (2.371) & -7.069*** (1.509) \\
\textbf{Leader Indicator} & -2.845*** (0.288) & -- & -1.645*** (0.242) &
-- & -2.053*** (0.511) & -- \\
\textbf{Observations} & 45,215 & 34,061 & 22,026 & 22,026 & 16,035 &
16,035 \\
\textbf{Athletes (FE)} & 15,762 & 12,181 & 10,075 & 10,075 & 8,408 &
8,408 \\
\textbf{Events (FE)} & 779 & 438 & 498 & 498 & 523 & 523 \\
\textbf{Clusters (FE)} & 149 & 178 & 178 & 178 & 178 & 178 \\
\textbf{RMSE} & 1.717 & 0.925 & 1.201 & 1.020 & 1.661 & 1.339 \\
\textbf{Adj.} \(\mathbf{R}^{\mathbf{2}}\) & -2.10 & 0.483 & -0.075 &
0.225 & -1.383 & -0.550 \\
\textbf{Within} \(\mathbf{R}^{\mathbf{2}}\) & -7.19 & -1.75 & -5.13 &
-3.42 & -13.20 & -8.21 \\
\textbf{First-stage F-stat} & 254.7 & 513.7 & 99.1 & 148.5 & 23.4 &
37.4 \\
\textbf{Wu-Hausman stat} & 1,901.0 & 918.2 & 509.1 & 508.6 & 298.5 &
297.7 \\
\textbf{Athlete FE} & \checkmark & \checkmark & \checkmark & \checkmark & \checkmark & \checkmark \\
\textbf{Event FE} & \checkmark & \checkmark & \checkmark & \checkmark & \checkmark & \checkmark \\
\textbf{Cluster FE} & \checkmark & \checkmark & \checkmark & \checkmark & \checkmark & \checkmark \\
\bottomrule()
\end{longtable}

\textbf{Notes:} \emph{p}~\textless{} 0.001 ***,~\emph{p}~\textless{}
0.01 **,~\emph{p}~\textless{} 0.05 *. 2SLS estimates of , where is
centered rank. The endogenous regressor is the theory-based drafting
benefit: for and 0 otherwise. Instrument: leave-one-out Group swim time.
Fixed effects for athlete, event, and cluster are included. Standard
errors clustered at the event level.\\
Leader coefficients are shown where included. Column labels correspond
to different subsamples and whether the leader dummy is included.
Drafting benefits are consistently significant and negative, with
stronger effects in smaller groups.

\section{Discussion and Robustness Results}
\emph{Instrument Strength\\
\strut \\
}\textbf{Table 12.} The causal effects of drafting position on log race
rank (2SLS Estimates)

\begin{longtable}[]{@{}
  >{\raggedright\arraybackslash}p{(\columnwidth - 10\tabcolsep) * \real{0.2080}}
  >{\raggedright\arraybackslash}p{(\columnwidth - 10\tabcolsep) * \real{0.1642}}
  >{\raggedright\arraybackslash}p{(\columnwidth - 10\tabcolsep) * \real{0.1498}}
  >{\raggedright\arraybackslash}p{(\columnwidth - 10\tabcolsep) * \real{0.1642}}
  >{\raggedright\arraybackslash}p{(\columnwidth - 10\tabcolsep) * \real{0.1642}}
  >{\raggedright\arraybackslash}p{(\columnwidth - 10\tabcolsep) * \real{0.1498}}@{}}
\toprule()
\multirow{2}{*}{\begin{minipage}[b]{\linewidth}\raggedright
\end{minipage}} &
\multicolumn{2}{>{\raggedright\arraybackslash}p{(\columnwidth - 10\tabcolsep) * \real{0.3139} + 2\tabcolsep}}{%
\begin{minipage}[b]{\linewidth}\raggedright
\textbf{Capped Drafting Position}
\end{minipage}} &
\multicolumn{3}{>{\raggedright\arraybackslash}p{(\columnwidth - 10\tabcolsep) * \real{0.4781} + 4\tabcolsep}@{}}{%
\begin{minipage}[b]{\linewidth}\raggedright
\textbf{Uncapped Drafting Position}
\end{minipage}} \\
& \begin{minipage}[b]{\linewidth}\raggedright
(1) Full

FE
\end{minipage} & \begin{minipage}[b]{\linewidth}\raggedright
(2) +Athlete FE
\end{minipage} & \begin{minipage}[b]{\linewidth}\raggedright
(3) +Event FE
\end{minipage} & \begin{minipage}[b]{\linewidth}\raggedright
(4) +Cluster FE
\end{minipage} & \begin{minipage}[b]{\linewidth}\raggedright
(5) +Athlete FE
\end{minipage} \\
\midrule()
\endhead
\textbf{Drafting Position (Fit)} & -0.504*** (0.069) & -4.741* (2.321) &
-0.046*** (0.011) & -0.083*** (0.013) & -0.011* (0.005) \\
\textbf{Leader Indicator} & -1.889*** (0.261) & -6.425* (3.159) &
-0.797*** (0.117) & -0.848*** (0.137) & -0.103* (0.048) \\
\textbf{Observations} & 23,544 & 9,316 & 34,061 & 34,061 & 34,061 \\
\textbf{RMSE} & 1.34 & 1.42 & 3.20 & 5.22 & 0.78 \\
\textbf{Adjusted} \(\mathbf{R}^{\mathbf{2}}\) & -0.30 & -1.43 & -2.90 &
-9.46 & 0.64 \\
\textbf{Within} \(\mathbf{R}^{\mathbf{2}}\) & -6.06 & -12.70 & -8.52 &
-24.80 & -0.94 \\
\textbf{First-stage F} & \textbf{87.3} & \textbf{8.42} & \textbf{400.3}
& \textbf{129.6} & \textbf{1,076.5} \\
\textbf{Wu-Hausman} & 530.9 & 93.0 & 3,755.8 & 3,518.6 & 1,038.0 \\
\textbf{Athlete Fixed Effects} & \checkmark & \checkmark & -- & -- & \checkmark \\
\textbf{Event Fixed Effects} & \checkmark & \checkmark & \checkmark & \checkmark & \checkmark \\
\textbf{Cluster Fixed Effects} & -- & -- & -- & \checkmark & \checkmark \\
\bottomrule()
\end{longtable}

\textbf{Notes:} \emph{p}~\textless{} 0.001 ***,~\emph{p}~\textless{}
0.01 **,~\emph{p}~\textless{} 0.05 *. All models are estimated via
two-stage least squares (2SLS), instrumenting Drafting Position using
the leave-one-out mean swim ability of other swimmers in the drafting
group. The dependent variable is (rank + 1 ), a log-transformation of
race placement. Columns (1)-(2) use a version of Drafting Position
capped at 5 to mitigate outlier influence; Columns (3)-(5) use the
uncapped measure. Event fixed effects are included in all models.
Athlete fixed effects appear in Columns (2) and (5); cluster fixed
effects in Columns (4) and (5). Standard errors are clustered at the
event level. \newline 

The coefficient on Leader Indicator reflects the performance association
of leading early. Negative estimates indicate a lower (better) log-rank
for lead swimmers, suggesting that leading is not empirically costly and
may instead reflect superior ability or small tactical and
congestion-related advantages, consistent with the interpretation in the
Theory-based specification section. First-stage F-statistics exceed the
conventional threshold of 10 in all but Column 2 (Table 12), confirming
instrument strength. Wu-Hausman tests reject exogeneity of Drafting
Position, supporting a causal interpretation. Under standard IV
assumptions (relevance, exclusion, monotonicity), the estimates identify
the Local Average Treatment Effect (LATE) of drafting position on
log-transformed race outcomes. The first-stage statistics generally
indicate relevant instruments, but the weaker values in some subsamples deserve
explicit caution. In particular, estimates with borderline first-stage
F-statistics should be treated as less precise and potentially more
vulnerable to weak-instrument bias. Large coefficients, especially in very small clusters, may be upwardly biased and should not be interpreted literally as exact causal magnitudes. Overall, the IV estimates support diminishing but meaningful drafting benefits most pronounced in compact groups. The leader coefficient is consistently negative, suggesting front positions carry small tactical or
congestion-related advantages rather than a net performance cost (Tables
6, 10--11). Supplementary OLS estimates using the within-group
drafter-position dummy approach, consistent with these IV results, are
reported Tables 13--15.

\textbf{Table 13.} Drafting position dummies FE model estimation results

\begin{longtable}[]{@{}
  >{\raggedright\arraybackslash}p{(\columnwidth - 8\tabcolsep) * \real{0.2655}}
  >{\raggedright\arraybackslash}p{(\columnwidth - 8\tabcolsep) * \real{0.1721}}
  >{\raggedright\arraybackslash}p{(\columnwidth - 8\tabcolsep) * \real{0.1914}}
  >{\raggedright\arraybackslash}p{(\columnwidth - 8\tabcolsep) * \real{0.1768}}
  >{\raggedright\arraybackslash}p{(\columnwidth - 8\tabcolsep) * \real{0.1941}}@{}}
\toprule()
\begin{minipage}[b]{\linewidth}\raggedright
\end{minipage} & \begin{minipage}[b]{\linewidth}\raggedright
\textbf{Estimate}
\end{minipage} & \begin{minipage}[b]{\linewidth}\raggedright
\textbf{Std. Error}
\end{minipage} & \begin{minipage}[b]{\linewidth}\raggedright
\textbf{t-value}
\end{minipage} & \begin{minipage}[b]{\linewidth}\raggedright
\[\mathbf{\Pr}\mathbf{( <}\mathbf{t}\mathbf{)}\]
\end{minipage} \\
\midrule()
\endhead
\textbf{First Drafter} & 71.9698 & 1.72555 & 41.70837 & 0.0000*** \\
\textbf{Second Drafter} & 39.4901 & 1.83705 & 21.49654 & 0.0000*** \\
\textbf{Third Drafter} & 28.1694 & 1.84785 & 15.24444 & 0.0000*** \\
\textbf{Fourth Drafter} & 23.5146 & 2.02665 & 11.60269 & 0.0000*** \\
\textbf{Fifth Drafter} & 19.5379 & 2.08455 & 9.37268 & 0.0000*** \\
\bottomrule()
\end{longtable}

\textbf{Notes}: p~\textless{} 0.001 ***,~p~\textless{} 0.01
**,~p~\textless{} 0.05 *. This table presents the results of an ordinary
least squares (OLS) estimation with fixed effects. The dependent
variable is Swim-Out Time (sec). The model includes fixed effects for
athlete\_id (29,194 levels) and event\_id (1,339 levels). Robust
standard errors are used to account for heteroskedasticity. The number
of observations is 168,391. The root mean squared error (RMSE) is 178.5.
The adjusted is 0.967166, and the within-group is 0.013724.

\textbf{Table 14.} Contd. drafting position dummies FE model estimation
results

\begin{longtable}[]{@{}
  >{\raggedright\arraybackslash}p{(\columnwidth - 8\tabcolsep) * \real{0.4118}}
  >{\raggedright\arraybackslash}p{(\columnwidth - 8\tabcolsep) * \real{0.1424}}
  >{\raggedright\arraybackslash}p{(\columnwidth - 8\tabcolsep) * \real{0.1584}}
  >{\raggedright\arraybackslash}p{(\columnwidth - 8\tabcolsep) * \real{0.1463}}
  >{\raggedright\arraybackslash}p{(\columnwidth - 8\tabcolsep) * \real{0.1410}}@{}}
\toprule()
\begin{minipage}[b]{\linewidth}\raggedright
\end{minipage} & \begin{minipage}[b]{\linewidth}\raggedright
\textbf{Estimate}
\end{minipage} & \begin{minipage}[b]{\linewidth}\raggedright
\textbf{Std. Error}
\end{minipage} & \begin{minipage}[b]{\linewidth}\raggedright
\textbf{t-value}
\end{minipage} & \begin{minipage}[b]{\linewidth}\raggedright
\[\mathbf{Pr( < t)}\]
\end{minipage} \\
\midrule()
\endhead
\multicolumn{5}{@{}>{\raggedright\arraybackslash}p{(\columnwidth - 8\tabcolsep) * \real{1.0000} + 8\tabcolsep}@{}}{%
\textbf{Panel A: OLS Estimation FE Model with Drafting Positions}} \\
\textbf{First Drafter} & 71.9698 & 1.72555 & 41.70837 & 0.00000 \\
\textbf{Second Drafter} & 39.4901 & 1.83705 & 21.49654 & 0.00000 \\
\textbf{Third Drafter} & 28.1694 & 1.84785 & 15.24444 & 0.00000 \\
\textbf{Fourth Drafter} & 23.5146 & 2.02665 & 11.60269 & 0.00000 \\
\textbf{Fifth Drafter} & 19.5379 & 2.08455 & 9.37268 & 0.00000 \\
\multicolumn{5}{@{}>{\raggedright\arraybackslash}p{(\columnwidth - 8\tabcolsep) * \real{1.0000} + 8\tabcolsep}@{}}{%
\textbf{Panel B: OLS Estimation FE Model with Drafting Positions and
Additional Controls}} \\
\textbf{Leader} & -26.7843 & 4.35217 & -6.15423 & 0.00000 \\
\textbf{Cluster (Swim Group)} & 0.65793 & 0.08622 & 7.63061 & 0.00000 \\
\textbf{Race Rank} & 0.39668 & 0.00723 & 54.85970 & 0.00000 \\
\textbf{First Drafter} & 48.5491 & 4.23790 & 11.45593 & 0.00000 \\
\textbf{Second Drafter} & 31.2118 & 1.79029 & 17.43395 & 0.00000 \\
\textbf{Third Drafter} & 23.5610 & 1.74702 & 13.48643 & 0.00000 \\
\textbf{Fourth Drafter} & 19.6363 & 1.89143 & 10.38174 & 0.00000 \\
\textbf{Fifth Drafter} & 15.4568 & 1.93864 & 7.97301 & 0.00000 \\
\textbf{Leader x Cluster (Swim Group)} & 1.29262 & 0.12643 & 10.22384 &
0.00000 \\
\bottomrule()
\end{longtable}
\textbf{Notes}: This table presents the results of two ordinary least
squares (OLS) estimations with fixed effects. The dependent variable is
Swim-Out Times. The models include fixed effects for athlete\_id (29,194
levels) and event\_id (1,339 levels). Robust standard errors are used to account
for heteroskedasticity. The number of observations is 168,391.\\ \newline
\textbf{Table 15.} Combined drafting position dummies FE model
estimation results

\begin{longtable}[]{@{}
  >{\raggedright\arraybackslash}p{(\columnwidth - 6\tabcolsep) * \real{0.2631}}
  >{\raggedright\arraybackslash}p{(\columnwidth - 6\tabcolsep) * \real{0.2456}}
  >{\raggedright\arraybackslash}p{(\columnwidth - 6\tabcolsep) * \real{0.2456}}
  >{\raggedright\arraybackslash}p{(\columnwidth - 6\tabcolsep) * \real{0.2456}}@{}}
\toprule()
\begin{minipage}[b]{\linewidth}\raggedright
\end{minipage} & \begin{minipage}[b]{\linewidth}\raggedright
\textbf{Model 1}
\end{minipage} & \begin{minipage}[b]{\linewidth}\raggedright
\textbf{Model 2}
\end{minipage} & \begin{minipage}[b]{\linewidth}\raggedright
\textbf{Model 3}
\end{minipage} \\
\midrule()
\endhead
\multirow{2}{*}{\textbf{Leader}} & -- & -26.7843*** & 0.8688 \\
& & (4.3522) & (3.9616) \\
\multirow{2}{*}{\textbf{Cluster (Swim Group)}} & -- & 0.6579*** & -- \\
& & (0.0862) & \\
\multirow{2}{*}{\textbf{Race Rank}} & -- & 0.3967*** & -- \\
& & (0.0072) & \\
\textbf{First Drafter} & 71.9698*** (1.7256) & 48.5491*** (4.2379) &
46.8571*** (4.3806) \\
\textbf{Second Drafter} & 39.4901*** (1.8371) & 31.2118*** (1.7903) &
26.0330*** (1.8769) \\
\textbf{Third Drafter} & 28.1694*** (1.8479) & 23.5610*** (1.7470) &
19.4259*** (1.8238) \\
\textbf{Fourth Drafter} & 23.5146*** (2.0267) & 19.6363*** (1.8914) &
17.4769*** (1.9906) \\
\textbf{Fifth Drafter} & 19.5379*** (2.0846) & 15.4568*** (1.9386) &
15.3214*** (2.0454) \\
\textbf{Observations} & 168,391 & 168,391 & 168,391 \\
\textbf{RMSE} & 178.5 & 169.2 & 175.0 \\
\textbf{Adj.} \(\mathbf{R}^{\mathbf{2}}\) & 0.9672 & 0.9705 & 0.9684 \\
\textbf{Within} \(\mathbf{R}^{\mathbf{2}}\) & 0.0137 & 0.1131 &
0.0050 \\
\bottomrule()
\end{longtable}

\textbf{Notes}: p~\textless{} 0.001 ***,~p~\textless{} 0.01
**,~p~\textless{} 0.05 *. This table presents the results of three
fixed-effects OLS estimations. The dependent variable is Swim-Out Times.
Models include fixed effects for athletes (29,194 levels) and events
(1,339 levels). Standard errors are estimated robust to
heteroskedasticity.

\emph{Positional Drafting Bandwagon Benefits Magnitude Interpretation}

The coefficient is interpreted as the semi-elasticity (For a detailed
discussion on accurately interpreting log-transformed variables in
regression models, see (Huntington-Klein, 2023) of adjusted rank with
respect to drafting disadvantage. Specifically, for binary , the
exponentiated coefficient gives the proportional change in adjusted
rank:

\[\%\text{~Change\ in\ Adjusted\ Rank~} \approx \left( e^{\beta} - 1 \right)\  \times \ 100\]

A negative thus implies that being in the latter drafting position group
leads to a statistically significant improvement in performance (i.e., a
lower adjusted rank). Early comparisons (e.g., 1-2 vs 3-4) yield large
and precisely estimated effects, consistent with the hypothesis of a
positional bandwagon drafting advantage. These effects saturate and at
least lose significance in later position\\
\strut \\
comparisons, suggesting a capped or diminishing marginal benefit of
drafting.\\
\strut \\
\textbf{Table 16}. Instrumental variable estimates: Effect of drafter
position on adjusted rank

\begin{longtable}[]{@{}
  >{\raggedright\arraybackslash}p{(\columnwidth - 12\tabcolsep) * \real{0.1558}}
  >{\raggedright\arraybackslash}p{(\columnwidth - 12\tabcolsep) * \real{0.2083}}
  >{\raggedright\arraybackslash}p{(\columnwidth - 12\tabcolsep) * \real{0.1341}}
  >{\raggedright\arraybackslash}p{(\columnwidth - 12\tabcolsep) * \real{0.1280}}
  >{\raggedright\arraybackslash}p{(\columnwidth - 12\tabcolsep) * \real{0.1266}}
  >{\raggedright\arraybackslash}p{(\columnwidth - 12\tabcolsep) * \real{0.1098}}
  >{\raggedright\arraybackslash}p{(\columnwidth - 12\tabcolsep) * \real{0.1374}}@{}}
\toprule()
\begin{minipage}[b]{\linewidth}\raggedright
\textbf{Comparison}
\end{minipage} & \begin{minipage}[b]{\linewidth}\raggedright
\textbf{Estimate (log pts)}
\end{minipage} & \begin{minipage}[b]{\linewidth}\raggedright
\textbf{Std. Error}
\end{minipage} & \begin{minipage}[b]{\linewidth}\raggedright
\textbf{Lower CI}
\end{minipage} & \begin{minipage}[b]{\linewidth}\raggedright
\textbf{Upper CI}
\end{minipage} & \begin{minipage}[b]{\linewidth}\raggedright
\textbf{P-value}
\end{minipage} & \begin{minipage}[b]{\linewidth}\raggedright
\textbf{\% Change}
\end{minipage} \\
\midrule()
\endhead
\textbf{1-2 vs 3-4} & -0.382 & 0.031 & -0.443 & -0.321 & 0.000 &
-31.8 \\
\textbf{2-3 vs 4-5} & -0.420 & 0.034 & -0.486 & -0.354 & 0.000 &
-34.3 \\
\textbf{3-4 vs 5-6} & -0.727 & 0.127 & -0.976 & -0.478 & 0.000 &
-51.9 \\
\textbf{4-5 vs 6-7} & -0.925 & 0.297 & -1.507 & -0.343 & 0.002 &
-60.3 \\
\textbf{5-6 vs 7-8} & -0.877 & 0.399 & -1.659 & -0.095 & 0.028 &
-58.1 \\
\textbf{6-7 vs 8-9} & -1.992 & 2.222 & -6.346 & 2.363 & 0.370 & -85.9 \\
\textbf{7-8 vs 9-10} & -2.067 & 2.510 & -6.987 & 2.854 & 0.401 &
-87.9 \\
\bottomrule()
\end{longtable}

Pooled Bandwagon Effects (Lower Rank Pool as Treatment on Rank Outcome)

\includegraphics[width=2.11129in,height=1.92191in]{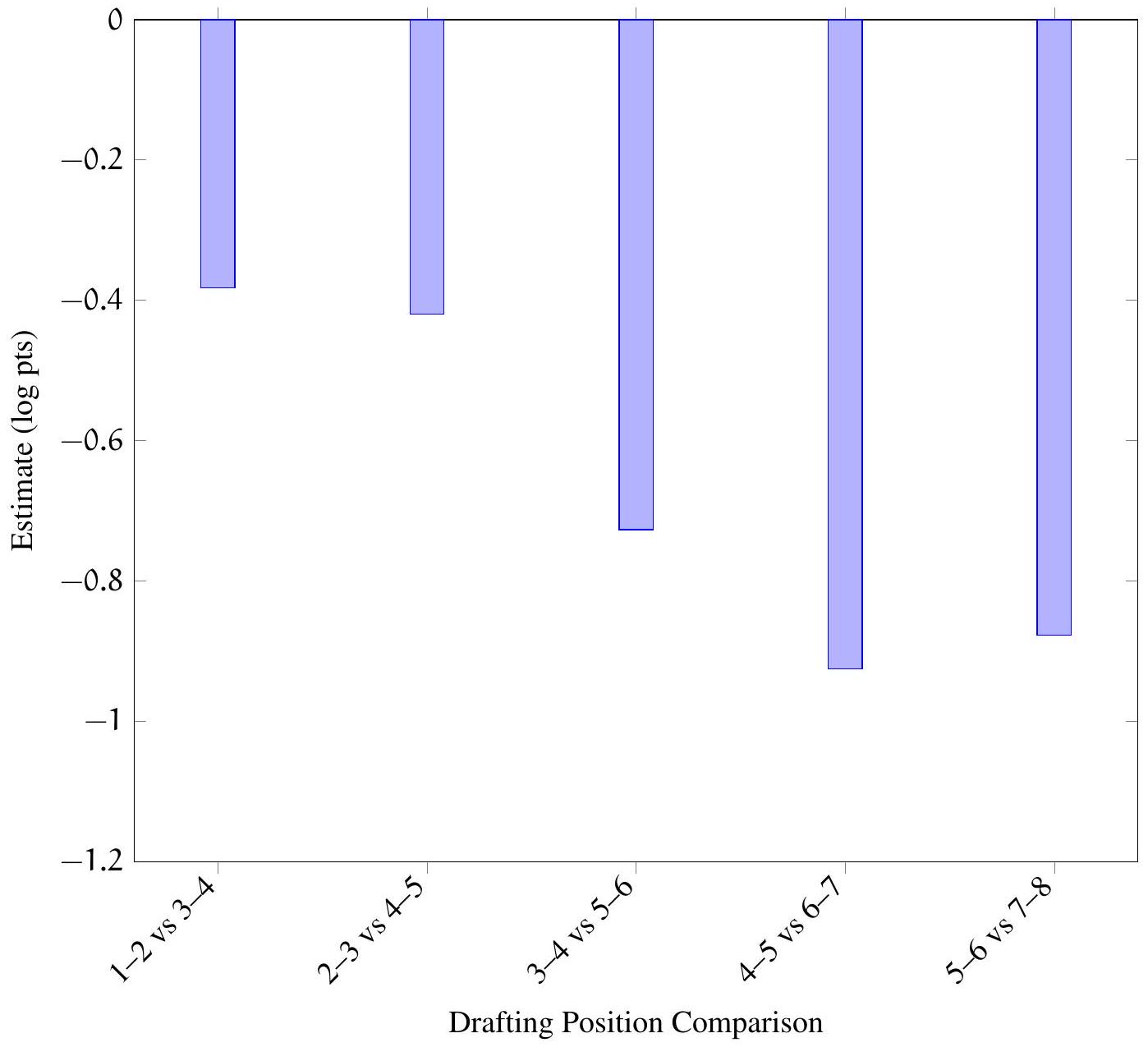}

\textbf{Figure 3}: Causal point estimates for the effect of latter
drafter position on overall better rank in log pts. (significant
comparisons only).

\section{Conclusion}

This paper provides a comprehensive analysis of strategic effort and
positional drafting benefits in finite multi-stage games with nonlinear
externalities, using causal evidence from drafting in endurance sports
competitions. Using panel data from triathlon races, log-linear models
of final-stage rank outcomes are estimated with endogenous drafting
position and a leave-one-out (LOO) measure of peer swim ability as an
instrument. Overall, the results support the view that drafting benefits
are strategic, nonlinear, and most pronounced in compact groups.

The empirical evidence indicates that deeper drafting positions are
associated with better performance, with stronger effects in smaller
groups. In very small clusters, the estimated magnitudes are especially
large, but these should be interpreted cautiously given weak-instrument
sensitivity in those subsamples. The most extreme percentage effects in
the smallest groups should be viewed as suggestive rather than
definitive, given the limited sample size and the possibility that weak
instruments result in unstable and potentially inflated point estimates.
The leader coefficient is consistently negative across specifications,
which may reflect tactical or congestion-related advantages of front
positions. However, an alternative explanation is residual selection:
stronger swimmers may systematically become leaders, and the IV strategy
may not completely eliminate this selection. The present evidence does
not definitively rule out this possibility, and both interpretations
should be weighed cautiously.

The pooled drafting positional bandwagon drafting benefits results reinforce the
broader pattern. Adjacent-position comparisons show that moving into
later drafting positions is often associated with better adjusted ranks,
but these gains flatten beyond the middle positions, consistent with
diminishing marginal returns and an interior optimum as predicted by the
theoretical model.

These findings relate to several strands of the literature. The
nonlinear, position-dependent structure of drafting benefits aligns with
theoretical work on strategic complementarities and supermodular games
(Topkis, 1998; Vives, 1990), where the value of an
agent\textquotesingle s action depends on the actions of others in a
non-additive manner. The identification strategy complements research on
peer effects and social interactions (Bramoullé et al., 2009; Angrist,
2014), demonstrating how institutional variation can help disentangle
true interaction effects from correlated baseline characteristics. The
results also extend empirical work on drafting in endurance sport
(Chatard \& Wilson, 2003; Delextrat et al., 2003; Bolon et al., 2023;
Reichel, 2025) by providing large-scale causal evidence from race data
rather than laboratory settings.

Several limitations should be acknowledged. First, drafting clusters are
inferred from swim-out times rather than directly observed, so some
measurement error and misclassification are unavoidable. Second,
unobserved environmental conditions such as water temperature, currents,
wind, or course layout may affect swim performance. Third, the COVID-19
start-policy changes that

generate identification may also have altered race dynamics beyond
drafting alone, including pacing and group formation. These caveats do
not overturn the main findings, but they do caution against
over-interpreting any single estimate as a fully structural effect. The
practical implications extend beyond triathlon. The findings suggest
that in any competitive environment with strategic positioning and
crowding externalities, agents benefit from understanding not only how
much effort to exert but also where to position themselves relative to
peers. For race organisers and coaches, the results highlight how start
procedures and group-formation policies can meaningfully alter
competitive dynamics and performance distributions. For policymakers,
the evidence demonstrates how regulatory changes---such as pandemic-era
restrictions---can serve as natural experiments that reveal otherwise
hidden strategic mechanisms.

More generally, the paper shows how strategic placement within a finite
competitive structure can shape outcomes when externalities are
nonlinear and position-dependent. Taken together, the findings
contribute to the literature on strategic externalities and multi-stage
games by showing that performance depends not only on effort, but also
on where an agent is positioned within a competitive pack. The evidence
is consistent with finite-horizon models in which the value of
positioning is real, nonlinear, and strongest when the competitive
environment is small enough for local interactions to matter most.

\textbf{Author Contributions}\\
F.R.: conception, design, supervision, materials, data processing,
analysis and interpretation, literature review, writing, critical
review.\newline
\textbf{\hfill\break
Declaration of Conflicting Interests}\\
The authors declared no potential conflicts of interest with respect to
the research, authorship, and/or publication of this article.\\ \newline
\textbf{Data Licensing and Availability}\\
Data are subject licensing and may be accessed under the
terms of the licensing agreement. Observations and data are via the Triathlon Statistics Austria website,
subject to the applicable license and permissions.

\end{document}